\documentclass[twocolumn]{aastex63}

\usepackage{tabularx}
\usepackage{url}
\usepackage{graphicx}
\usepackage[T1]{fontenc}
\usepackage{ae,aecompl}
\usepackage{booktabs}
\usepackage{natbib}
\usepackage{multirow}
\usepackage{multirow}
\usepackage{appendix}
\DeclareGraphicsExtensions{.pdf,.png,.jpg,.mps,.eps,.ps}
\usepackage{blindtext}
\usepackage{longtable}
\usepackage{tabu}
\usepackage{lineno}

\usepackage{chngcntr}

\def\unit #1{\,{\rm #1}}

\newcommand\kms{\rm \,\unit{km\,s^{-1}}}

\newcommand\cmsqi{\rm \,\unit{cm^{-2}}}

\newcommand\kev{\rm \,\unit{keV}}

\newcommand\funit{\rm \,erg\,cm^{-2}\,s^{-1}}

\newcommand\funita{\rm\,erg\,cm^{-2}\,s^{-1}\angstrom^{-1}}

\newcommand\pc{\unit{pc}}

\newcommand\mpc{\unit{Mpc}}

\newcommand\ev{\unit{\, eV}}

\newcommand{\angstrom}{\mbox{\normalfont\AA}}

\newcommand\swift{{\it Swift}}
\newcommand\xmm{{\it XMM-Newton}}

\newcommand\nicer{{\it NICER}}

\newcommand{\angs}{\mbox{\normalfont\AA}}

\def\aap{A\&A} \def\apjl{ApJ}  
 \def\apjs{ApJS}

 \submitjournal{ApJ}

\shorttitle{Changing look AGN 1ES 1927+654}

\begin{document}

\title{ A radio, optical, UV and X-ray view of the enigmatic changing look Active Galactic Nucleus 1ES~1927+654 from its pre- to post-flare states. }

\author[0000-0003-2714-0487]{Sibasish Laha} 

\affiliation{Center for Space Science and Technology, University of Maryland Baltimore County, 1000 Hilltop Circle, Baltimore, MD 21250, USA.}
\affiliation{Astrophysics Science Division, NASA Goddard Space Flight Center,Greenbelt, MD 20771, USA.}
\affiliation{Center for Research and Exploration in Space Science and Technology, NASA/GSFC, Greenbelt, Maryland 20771, USA}

\author[0000-0000-0000-0000]{Eileen Meyer}
\affiliation{Department of Physics, University of Maryland, Baltimore County, 1000 Hilltop Circle, Baltimore, MD 21250, USA.}

\author[0000-0003-1101-8436]{Agniva Roychowdhury}
\affiliation{Department of Physics, University of Maryland, Baltimore County, 1000 Hilltop Circle, Baltimore, MD 21250, USA.}

\author[0000-0000-0000-0000]{Josefa Gonz\'alez ~Becerra} 
\affiliation{Instituto de Astrof\'isica de Canarias (IAC), E-38200 La Laguna, Tenerife, Spain}
\affiliation{Universidad de La Laguna (ULL), Departamento de Astrof\'isica, E-38206 La Laguna, Tenerife, Spain}

\author[0000-0000-0000-0000]{J.~A.~Acosta--Pulido}
\affiliation{Instituto de Astrof\'isica de Canarias (IAC), E-38200 La Laguna, Tenerife, Spain}
\affiliation{Universidad de La Laguna (ULL), Departamento de Astrof\'isica, E-38206 La Laguna, Tenerife, Spain}

\author[0000-0002-8843-9581]{Aditya Thapa}
\affiliation{Department of Physics, University of Maryland, Baltimore County, 1000 Hilltop Circle, Baltimore, MD 21250, USA.}

\author[0000-0003-4790-2653]{Ritesh Ghosh}
\affiliation{Inter-University Centre for Astronomy and Astrophysics (IUCAA), Pune, 411007, India..}

\author[0000-0000-0000-0000]{Ehud Behar}
\affiliation{Department of Physics, Technion, Haifa 32000, Israel}

\author[0000-0000-0000-0000]{Luigi C. Gallo}
\affiliation{Department of Astronomy \& Physics, Saint Mary’s University, 923 Robie Street, Halifax, Nova Scotia, B3H 3C3, Canada}

\author[0000-0002-2180-8266]{Gerard A.\ Kriss}
\affiliation{Space Telescope Science Institute, 3700 San Martin Drive, Baltimore, MD 21218, USA}

\author[0000-0003-0543-3617]{Francesca Panessa}
\affiliation{INAF - Istituto di Astrofisica e Planetologia Spaziali, via Fosso del Cavaliere 100, I-00133 Roma, Italy}

\author[0000-0002-4622-4240]{Stefano Bianchi}
\affiliation{Dipartimento di Matematica e Fisica, Universit\`a degli Studi Roma Tre, Via della Vasca Navale 84, I-00146, Roma, Italy}

\author[0000-0002-1239-2721]{Fabio La Franca}
\affiliation{Dipartimento di Matematica e Fisica, Universit\`a degli Studi Roma Tre, Via della Vasca Navale 84, I-00146, Roma, Italy}

\author[0000-0002-1239-2721]{Nicolas Scepi}
\affiliation{JILA, University of Colorado and National Institute of Standards and Technology, 440 UCB, Boulder, CO 80309-0440, USA.}

\author[0000-0003-0936-8488]{Mitchell C.~Begelman}
\affiliation{JILA, University of Colorado and National Institute of Standards and Technology, 440 UCB, Boulder, CO 80309-0440, USA.}


\author[0000-0002-1239-2721]{Anna Lia Longinotti}
\affiliation{Instituto de Astronomía, Universidad Nacional Autónoma de México, Circuito Exterior, Ciudad Universitaria, Ciudad de México 04510, México}

\author[0000-0003-0083-1157]{Elisabeta Lusso}
\affiliation{Dipartimento di Fisica e Astronomia, Università di Firenze, Via G. Sansone 1, 50019, Sesto Fiorentino, Firenze, Italy}
\affiliation{INAF - Osservatorio Astrofisico di Arcetri, Largo Enrico Fermi 5, 50125, Firenze, Italy}

\author[0000-0000-0000-0000]{Samantha Oates}

\affiliation{Birmingham Institute for Gravitational Wave Astronomy and School of Physics and Astronomy, University of Birmingham, Birmingham B15 2TT, UK}

\author[0000-0002-2555-3192]{Matt Nicholl}

\affiliation{Birmingham Institute for Gravitational Wave Astronomy and School of Physics and Astronomy, University of Birmingham, Birmingham B15 2TT, UK}

\author[0000-0003-1673-970X]{S. Bradley Cenko}
\affiliation{Astrophysics Science Division, NASA Goddard Space Flight Center,Greenbelt, MD 20771, USA.}
\affiliation{Joint Space-Science Institute, University of Maryland, College Park, MD 20742, USA}

\correspondingauthor{Sibasish Laha}
\email{sibasish.laha@nasa.gov,sib.laha@gmail.com}


\begin{abstract}
	
	The nearby type-II AGN 1ES1927+654 went through a violent changing-look (CL) event beginning December 2017 during which the optical and UV fluxes increased by four magnitudes over a few months, and broad emission lines newly appeared in the optical/UV. By July 2018 the X-ray coronal emission had completely vanished, only to reappear a few months later. In this work we report the evolution of the radio, optical, UV and X-rays from the pre-flare state through mid-2021 with new and archival data from the {\it Very Long Baseline Array}, the {\it European VLBI Network}, the {\it Very Large Array}, the {\it Telescopio Nazionale Galileo}, {\it Gran Telescopio Canarias}, {\it The Neil Gehrels Swift observatory} and \xmm{}. The main results from our work are: \emph{(i)} The source has returned to its pre-CL state in optical, UV, and X-ray; the disk--corona relation has been re-established as has been in the pre-CL state, with an $\alpha_{\rm OX}\sim 1.02$. The optical spectra are dominated by narrow emission lines. \emph{(ii)} The UV light curve follows a shallower slope of $\propto t^{-0.91\pm 0.04}$ compared to that predicted by a tidal disruption event. We conjecture that a magnetic flux inversion event is the possible cause for this enigmatic event. \emph{(iii)} The compact radio emission which we tracked in the pre-CL (2014), during CL (2018) and post-CL(2021) at spatial scales $<1\pc$ was at its lowest level during the changing look event in 2018, nearly contemporaneous with a low $2-10\kev$ emission. The radio to X-ray ratio of the compact source $L_{\rm Radio}/L_{\rm X-ray}\sim 10^{-5.5}$, follows the Gudel-Benz relation, typically found in coronally active stars, and several AGN. \emph{(iv)} We do not detect any presence of nascent jets at the spatial scales of $\sim 5-10\pc$.

\end{abstract}


\vspace{0.5cm}


\section{INTRODUCTION}

The exact geometry and functioning of the central engines of active galactic nuclei (AGN) are still highly debated. The long duty cycle of AGN \citep[$ 10^7-10^9$ years][]{Marconi2004,2015MNRAS.451.2517S} compared to human timescales 
is expected to prevent direct observation of the ignition or quenching of an AGN. 
However, recent discoveries of so-called changing-look AGN (CL-AGN), have given us rare glimpses of extreme changes in the AGN state in a few months to years. Not all of these changes happen in the same way, which intimates the complexity of the physical mechanisms at work in the central engine.

CL-AGN are rare, with only a few dozen candidates in the literature. 
The term applies both to sources which change from an approximately type I to type II state and vice versa. For example, one of the earliest discovered CL-AGN, Mkn 1018, transitioned from a Seyfert 1.9 to type 1 over the course of $\sim5$ years in the early 1980s \citep{Cohen1986}. The higher activity state was maintained
for decades before significantly dimming (factor of $\sim 25$) and changing back to Seyfert type 1.9
during 2013-2015. The X-ray spectra showed no detectable absorption, hence the dramatic change
must be intrinsic to the accretion disk emission itself, suggesting major changes in the accretion
flow \citep{Husemann2016}. 

In contrast, in Mrk 590 (one of the best-observed CL-AGN) the initially type 1 source dimmed in the optical by a factor of $\sim 100$ over three decades, with complete disappearance of the formerly strong and broad H$\alpha$ emission line \citep{Denney2014,Mathur2018}. A prominent soft X-ray excess during the ``bright state", when the broad H$\alpha$ and H$\beta$ lines were present in the optical band, could not be explained by disk reflection. No
obscuration in X-rays was detected, but ultra-fast outflows and a nascent jet recently discovered
with VLBI are present \citep{yang21}. It has been suggested that the CL nature of this AGN could be due to
episodic accretion events, as it has been observed to re-brighten and dim more than once. A similar case is Mrk 335; originally one of the X-ray-brightest AGN, the flux dropped dramatically in 2007
\citep{Grupe2012}. Since then, optical, UV and X-ray monitoring suggest the corona has ‘collapsed’ in toward the black hole \citep{Gallo2018, Tripathi2020}, and the source sometimes forms a collimated outflow in X-ray flare states \citep{Wilkins2015a, Wilkins2015b, Gallo2019}.

The nearby ($z=0.017$, luminosity distance= $74.2\mpc$) CL-AGN 1ES~1927+654, the subject of this paper, is a more recent discovery (RA=$291.83$, dec=$65.56$ in degrees, J$2000$). It had been previously classified as a true type-II AGN defying the unification model \citep{panessa02,bianchi12}, because there had been no detection of broad H$\alpha$ and H$\beta$ emission lines, neither was there any line-of-sight obscuration in the optical, UV or X-rays \citep[][and references therein]{Boller2003,Gallo2013}. \cite{Tran2011} suggested that the broad line region (BLR) is absent due to low Eddington ratio of this source (i.e., there not being enough continuum emission to light up the BLR). \cite{Wang2012} suggested that the AGN in 1ES~1927+654 is young and did not have the time to create a BLR.

The dramatic CL event in 1ES~1927+654 began with a significant rise in the optical/UV in Dec 2017 \citep[detected by the ATLAS survey,][]{trak19}. It continued to rise in luminosity for $\sim 150$ days, and by the peak in March 2018 the optical/UV had increased by four magnitudes (almost a factor of 100). Afterwards the optical/UV decayed with a $t^{-5/3}$ TDE-like light-curve \citep{trak19}. The optical spectrum just after the flare began was dominated by a blue continuum and several narrow emission lines including H$\alpha$, H$\beta$, O[III] $\lambda5007$, implying that the broad line region had not yet responded. The narrow emission lines were consistent with gas photo-ionized by AGN continuum \citep{trak19}.  Strong broad H$\alpha$ and H$\beta$ emission lines (FWHM$\sim 17,000\kms$) started to appear around $\sim 100$ days after the flare (between March 6th and April 23 in 2018). The lines remained strong for the next $\sim 300$ days, after which there was a large Balmer decrement, indicating the presence of dust absorption. The X-ray monitoring of the source started after $\sim 100$ days of the initial flare, after which the X-ray emission started decreasing rapidly and reached a minimum of $10^{-3}$ times its original flux in about $\sim 200$ days after the initial flare. In its lowest flux state, the spectrum shows only soft ($0.3-2 \kev$) emission with no signature from $2-10\kev$, as expected for coronal emission. This indicates that the X-ray emitting corona was completely destroyed in the process \citep{Ricci2021}. The X-ray spectrum soon recovered to a flux level of $\sim 10$ times that of the pre-flare flux in another 100 days (i.e., by April 2019).

In this work we investigate the present active state of 1ES~1927+654, with new observations obtained through 2021. We also utilize archival observations (in all wavelength bands) to better trace the full evolution of this source before, during, and after the CL event. In this paper we address in particular the following questions: \\
1. What is the cause for this violent event? Is it due to the changes in external rate of mass supply (accretion efficiency due to a TDE), or internal mechanism related to the change of polarity of the magnetic field of the accretion disk?\\
2. In the current post-flare state, is the X-ray corona fully formed?\\
3. Is the disk-corona relation established? \\
4. What is the origin of the soft X-ray emission, which was still sustained when the X-ray corona completely vanished? \\
5. Is this really a true type-II AGN? What do we infer from the broad line emission region detection?  \\
6. How has the core ($<1\pc$) radio luminosity evolved during the entire cycle from pre- to post-flare states? \\
7. Are there any indications of nascent jet formation or winds, three years after the flare erupted?

The paper is arranged as follows: Section \ref{sec:obs} describes the observation, data reduction techniques, and data analysis of the multi-wavelength observations. Section \ref{sec:results} describes the main results from our observational campaigns.
This is followed by discussions in Section \ref{sec:discussion} and conclusions in Section \ref{sec:conclusions}. Throughout this paper, we assumed a cosmology with $H_{0} = 71\kms \mpc^{-1}, \Omega_{\Lambda} = 0.73$ and $\Omega_{M} = 0.27$.


\begin{table*}

\centering
  \caption{The multiwavelength observations of 1ES~1927+654. \label{Table:obs}}
  \begin{tabular}{cccccccc} \hline\hline 

Observation band	& Telescopes			&observation date	&observation ID	& Net exposure & Short-id	\\
			        &   				    & YYYY-MM-DD               &		        &(Sec)	\\ \hline 

X-ray and UV		&{\it Swift-XRT/UVOT}&2018-05-17	& 00010682001	&2190   &S01	\\

''		            & ''                 &2018-05-31	& 00010682002	    &1781   &S02	\\
''		            & ''                 &2018-06-14 	& 00010682003	&2126   &S03	\\
''		            & ''                 &2018-07-10	& 00010682004	    &1599   &S04	\\
''		            & ''                 &2018-07-24	& 00010682005	    &2302   &S05	\\
''		            & ''                 &2018-08-07	& 00010682006	    &2171   &S06	\\
''		            & ''                 &2018-08-23 & 00010682007	    &1977   &S07	\\
''		            & ''                 &2018-10-03		& 00010682008	&1252   &S08	\\
''		            & ''                 &2018-10-19 	& 00010682009	&966   &S09	\\
''		            & ''                 &2018-10-23		& 00010682010	&1591   &S10	\\
''		            & ''                 &2018-11-21		& 00010682011	&2174   &S11	\\
''		            & ''                 &2018-12-06 	& 00010682012	&1568   &S12	\\
''		            & ''                 &2018-12-12		& 00010682013	&1986   &S13	\\
''		            & ''                 &2019-03-28		& 00010682014	&2138   &S14	\\
''		            & ''                 &2019-11-02		& 00088914001	&207   &S14A	\\

''                   &''                  &2021-02-24	& 00010682015  &864	    &S15	\\
 ''   		        & ''                   &2021-03-09	& 00010682017	&308    &S17	\\
''		            &   ''                &2021-03-10	& 00010682018	&1004   &S18	\\
''		            & ''                 &2021-03-11		& 00010682019	&1064   &S19	\\
''		            &   ''               &2021-03-12		& 00010682020	&919    &S20	\\
''		            & ''                 &2021-03-13		& 00010682021	&894    &S21	\\
''		            & ''                 &2021-04-12		& 00010682023	&1900   &S23	\\
''		            & ''                 &2021-05-18		& 00010682025	&710   &S25	\\
''		            & ''                 &2021-06-17		& 00010682026	&1513   &S26	\\
''		            & ''                 &2021-07-15		& 00010682027	&527   &S27	\\
''		            & ''                 &2021-08-20		& 00010682028	&1376   &S28	\\
''		            & ''                 &2021-10-20		& 00010682029	&1696   &S29	\\
''		            & ''                 &2021-11-20		& 00010682030	&1556   &S30	\\
''		            & ''                 &2021-12-20		& 00010682031	&1631   &S31	\\
\hline

''		            & \xmm{} EPIC-pn/OM  &2011-05-20		& 0671860201	&28649   & X1	\\

\hline

Optical     &{\it TNG}  & 2011-06-02    &   -           & 1800\\
''		&{\it GTC}			&	2021-03-10	& GTC2021-176	& 450	\\
''			&''			&	2021-05-04	& ''	& 450	\\
\hline


Radio			&{\it VLA}		&	1992-01-31	& AS0452 & 210 	\\
''			& ''		&	1998-06-06	& AB0878 & 240	\\
''			&{\it VLBI}		&	2013-08-10	& EG079A & 5400		\\
''			& ''		    &			2014-03-25	& EG079B & 5400    \\
''			& ''			&			2018-12-04	& RSY07  & 10800 \\
''			& ''			&			2021-03-15	& 21A-403 & 12600 \\

\hline 
\end{tabular}  

{{\it TNG} = Telescopio Nazionale Galileo, {\it GTC} = Gran Telescopio CANARIAS , {\it VLA} = Very Large Array, {\it VLBI}= Very Large Baseline Interferometer. }
\end{table*}


\begin{table*}
\fontsize{8}{10}
\centering
  \caption{The spectral parameters obtained using {\it Swift} and \xmm{} UV and X-ray observations of 1ES~1927+654.  \label{Table:xray_obs}} 
  \begin{tabular}{cccccccccccc} \hline\hline 

ID(MM/YY)			&$F_{0.3-2\kev}^{\rm A}$	&$F_{2-10\kev}^{\rm A}$ &$F_{1.5-2.5\kev}^{\rm A}$  & kT& $\Gamma$ & UV filter& UV flux density$^{\rm B}$ &$\alpha_{\rm OX}$ & $\rm \chi^2/\chi^2_{\nu}$\\ 
				    &        	                &                       &                           & (keV)       &&		&	\\ \hline 

X1 (05/11)	    &   $9.41\pm0.66$   & $3.92\pm0.08$ & $1.64\pm0.02$ &$0.20\pm0.01$               & $2.21_{-0.02}^{+0.02}$ &UVM2	& $1.34\pm0.03$ &1.004 & $185/1.37$  \\ \hline

S01 (05/18)     &   $26.41\pm2.12$   &$0.06\pm0.02$      & $0.25\pm0.04$     &$0.15\pm0.01$	&$4.94_{-0.97}^{+2.48}$	&UVW2   &$16.17\pm0.72$	    &1.734 & $45/1.02$\\
S02 (05/18)     &   $8.32\pm1.33$   &$0.04\pm0.02$      & $0.06\pm0.02$     &$-$	        &$4.86_{-0.30}^{+0.33}$	&UVW2   &$14.67\pm0.67$	    &1.955 & $7.10/1.01$\\
S03 (06/18)     &   $4.21\pm0.71$   &$<0.36$            & $0.18\pm0.14$     &$-$	        &$3.61_{-0.60}^{+0.62}$	&UVW2   &$13.00\pm0.59$	    &1.752 & $1.76/0.59$\\
S04 (07/18)     &   $-$             &$-$                & $-$	            &$-$	        &$-$                    &UVW2   &$11.30\pm0.54$	    &- & $-$\\
S05 (07/18)     &   $<0.393$        &$<0.14$            & $<0.10$           &$-$	        &$-$	                &UVW2   &$10.74\pm0.50$	    &- & $-$\\
S06 (08/18)     &   $<1.44$         &$<0.01$            & $<0.01$           &$-$	        &$-$	                &UVW2   &$10.57\pm0.48$	    &- & $-$\\
S07 (08/18)     &   $5.01\pm1.12$    &$<0.12$            & $0.09\pm0.07$     &$-$           &$4.26_{-0.60}^{+0.63}$ &UVW2   &$9.57\pm0.44$	    &1.816 & $3.47/0.87$\\
S08 (10/18)     &   $24.14\pm1.66$   &$7.17\pm3.15$      & $1.49\pm0.24$     &$0.14\pm0.01$	&$1.40_{-1.08}^{+0.87}$	&UVW2   &$5.67\pm0.31$	    &1.261 & $26.49/0.80$ \\
S09 (10/18)     &   $44.63\pm2.32$   &$8.57\pm2.00$      & $4.92\pm0.42$     &$0.18\pm0.03$	&$2.62_{-0.23}^{+0.19}$	&UVW2   &$9.00\pm0.48$	    &1.139 & $51.14/0.95$\\
S10 (10/18)     &   $33.10\pm1.75$   &$2.85\pm1.09$      & $1.72\pm0.30$     &$0.14\pm0.01$	&$2.73_{-0.79}^{+0.43}$	&UVW2   &$8.50\pm0.43$  	&1.305 & $33.89/0.68$\\
S11 (11/18)     &   $32.31\pm1.92$   &$1.05\pm0.54$      & $1.02\pm0.20$     &$0.15\pm0.01$	&$3.29_{-0.63}^{+0.43}$	&UVW2   &$7.96\pm0.37$  	&1.381 & $32.66/0.76$\\
S12 (12/18)     &   $19.21\pm1.71$   &$0.89\pm0.27$      & $0.85\pm0.18$     &$-$	        &$3.60_{-0.63}^{+0.43}$	&UVW2   &$8.52\pm0.43$	    &1.423 & $25.30/1.15$\\
S13 (12/18)     &   $40.61\pm1.98$   &$3.28\pm0.83$      & $2.17\pm0.21$     &$0.16\pm0.01$	&$2.70_{-0.35}^{+0.27}$	&UVW2   &$7.95\pm0.37$	    &1.255 & $78.50/1.38$\\ 
S14 (03/19)     &   $51.22\pm1.86$   &$9.81\pm1.35$      & $5.41\pm0.35$     &$0.18\pm0.01$	&$2.46_{-0.16}^{+0.14}$	&UVW2   &$6.09\pm0.30$	    &1.058 & $137.88/1.14$\\
S14A (11/19)    &   $68.91\pm7.88$   &$27.91\pm6.27$     & $12.35\pm1.52$    &$-$	        &$2.42_{-0.15}^{+0.15}$	&UVW2   &$4.39\pm0.30$	    &0.866 & $25.67/1.35$\\

S15 (02/21)     &   $19.12\pm2.22$   &$3.46\pm1.22$  &$2.12\pm0.28$  &	$0.16\pm 0.02$	&$2.59_{-0.51}^{+0.36}$	&UVW2   &$2.17\pm0.13$	&1.042    & $20/0.83$ \\
S17 (03/21)     &   $32.31\pm5.21$   &$1.34\pm0.72$  &$2.01\pm0.45$  &	$0.12\pm 0.02$	&$1.79_{-0.91}^{+0.78}$	&UVW2	&$2.17\pm0.19$  &1.051    & $10/1$ \\
S18 (03/21)     &   $24.55\pm2.45$   &$4.78\pm0.22$  &$3.01\pm0.19$  &	$0.19\pm 0.04$	&$2.68_{-0.25}^{+0.23}$	&UVW2	&$2.28\pm0.13$  &0.992    & $30/0.88$ \\
S19 (03/21)     &   $25.73\pm1.86$   &$5.37\pm1.39$  &$3.02\pm0.23$  &	$0.20\pm 0.04$	&$2.52_{-0.24}^{+0.28}$	&UVW2	&$2.30\pm0.13$  &0.993    & $55/1.34$ \\	
S20 (03/21)     &   $20.44\pm3.21$   &$6.02\pm2.82$  &$2.04\pm0.24$  &	$0.13\pm 0.02$	&$2.09_{-0.87}^{+0.76}$	&UVW2	&$2.17\pm0.13$  &1.049    & $25/1.19$ \\
S21 (03/21)     &   $32.31\pm4.23$   &$5.24\pm1.33$  &$4.10\pm0.19$  &	$0.28\pm 0.09$	&$2.84_{-0.26}^{+0.26}$	&UVW2	&$2.04\pm0.11$  &0.922    & $44/1.13$ \\
S23 (04/21)     &   $20.24\pm1.12$   &$6.32\pm0.71$  &$3.12\pm0.21$  &	$0.16\pm 0.03$	&$2.35_{-0.18}^{+0.16}$	&UVM2	&$1.94\pm0.07$  &0.959 & $60/1.15$\\

S25 (05/21)	    &   $23.91\pm2.91$   & $5.23\pm1.36$ & $2.97\pm0.44$ & $-$               &$2.76_{-0.18}^{+0.19}$ &UVM2	& $1.83\pm0.09$ &0.958 & $15.70/0.98$\\
S26 (06/21)	    &   $18.21\pm1.52$   & $6.64\pm1.43$ & $2.88\pm0.29$ & $0.18\pm0.05$     &$2.34_{-0.31}^{+0.22}$ &UVM2	& $1.94\pm0.09$ &0.972 & $57.17/1.43$\\

S27 (07/21)	    &   $19.15\pm2.41$   &$6.09\pm 2.75$ & $3.01\pm0.64$ & $0.58\pm0.03$     &$1.62_{-0.88}^{+0.71}$ &UVM2	& $1.94\pm0.11$ &0.965 & $18.04/1.64$\\

S28 (08/21)	    &   $17.22\pm1.45$   & $5.90\pm1.02$ & $2.78\pm0.28$ & $-$               &$2.50_{-0.12}^{+0.12}$ &UVW2	& $2.09\pm0.09$ &0.991 & $20.46/0.76$\\

S29 (10/21)	    &   $9.74\pm0.85$   & $4.42\pm0.81$ & $1.86\pm0.18$ & $-$               &$2.35_{-0.12}^{+0.12}$ &UVW2	& $2.06\pm0.09$ &1.055 & $17.68/0.68$\\

S30 (11/21)    &   $14.04\pm1.13$   & $4.75\pm0.84$ & $2.27\pm0.23$ & $-$               &$2.52_{-0.12}^{+0.12}$ &UVW2	& $2.04\pm0.09$ &$1.022$ & $23.69/0.82$\\

S31 (12/21)	    &   $20.07\pm1.24$   & $5.94\pm0.81$ & $2.99\pm0.24$ & $-$               &$2.59_{-0.09}^{+0.09}$ &UVW2	& $2.06\pm0.09$ &$0.981$ & $47.33/1.01$\\

\hline 
\end{tabular}  
$^{\rm A}$ Flux in units of $ 10^{-12}\funit{}$\\
$^{\rm B}$ UV flux density in units of $ 10^{-15}\funita{}$\\
$\alpha_{\rm OX}=-0.385\log(\rm F_{2\kev}/F_{2500\rm \AA})$\\
The UV flux density were corrected for Galactic absorption using the correction magnitude of $\rm A_{\lambda}=0.690$ obtained from NED.

\end{table*}


\section{Observation, data reduction and data analysis}\label{sec:obs}

\subsection{Swift XRT and UVOT}

New observations of 1ES~1927+654 were carried out by {\it The Neil Gehrels Swift Observatory} (from now on \swift{}) during 2021 at a monthly cadence under a Director's Discretionary Time (DDT) program (PI: S.Laha), which we present here. We have also analyzed all the archival \swift{} observations from 2018 and 2019 to make a comparison between the flaring and post-flare states. Table \ref{Table:obs} lists all the \swift{} observations, and their short ids (S01-S29). The \swift{} X-ray Telescope \citep[XRT ][]{Burrows2005} observations were performed mostly in photon counting mode, and a few times in window timing mode. We analysed the XRT data with standard procedures using \texttt{XRTPIPELINE}. The HEASOFT package version 6.28 and most recent calibration data base (CALDB) were used for filtering, and screening of the data. In the cases taken in photon-counting mode, the source regions were selected using $40''$ circles centered around the centroid of the source, and the background regions were selected with similarly sized circles away from the source. In the observations that were taken in window timing mode, the source and background regions were selected in boxes 40 pixels long. We use the standard grade selections of 0–2 for the window timing mode. Source photons for the light curve and spectra were extracted with \texttt{XSELECT} in both modes. The auxiliary response files (ARFs) were created using the task \texttt{xrtmkarf} and using the response matrices obtained from the latest Swift CALDB. We bin the data using \texttt{grppha} to have at least 20 counts per bin.

The Ultraviolet-Optical Telescope \citep[UVOT ][]{Roming2005} observed the source 1ES~1927 in 2018-2019 with all the six filters i.e. in the optical (V, B, U) bands and the near UV (W1, M2, W2) bands, but only with UVM2 and UVW2 in 2021. Since we are interested in a consistent photometric data point in the UV, for comparison over time, we choose to use UVW2 for all  observations, and UVM2 where UVW2 is not present. We used the standard UVOT reprocessing methods and calibration database \citep{breevald2011} to obtain the monochromatic flux density in the UV (with a 5$''$ selection radius) and the corresponding statistical and systematic errors were obtained by the \texttt{uvotsource} task. The UV flux densities were corrected for Galactic absorption using the correction magnitude of $\rm A_{\lambda}=0.690$ obtained from NASA Extragalactic Database (NED \footnote{https://ned.ipac.caltech.edu}).

\subsubsection{Swift XRT spectral analysis}
To fit the Swift XRT $0.3-10\kev$ spectra, we assumed a simple baseline model of {\tt tbabs*(bbody+powerlaw)}, following the pre-flare 2011 spectral modeling \citep{Gallo2013}. The {\tt tbabs} model represents the neutral Galactic absorption, {\tt bbody} model describes the soft X-ray excess and the {\tt powerlaw} model describes the Inverse-Compton emission from the AGN corona. The poor signal to noise data and the low flux state of the source in most observations did not allow us to use more complex models. See Table \ref{Table:xray_obs} for details of the best fit parameters and the fit statistics ($\chi^2/\chi^2_{\nu}$) for every observation.  We note that in the observations S04, S05, and S06 we do not detect any X-ray photons with XRT, indicating an X-ray low flux state. In the cases of S04 and S05 we could put upper limits on the fluxes. As we see from the fit statistics in Table \ref{Table:xray_obs}, in most cases the baseline model gives a satisfactory fit in the $0.3-10\kev$ band. We also note from Table \ref{Table:xray_obs} that the powerlaw slope has been very steep during the changing look phase ($\Gamma \sim 5 $) which gradually reached its pre-flare value of $\Gamma=2.21$, over a period of $\sim 1200$ days. We however, do not have any \swift{} monitoring data between Dec-2019 and Feb-2021 and so cannot comment anything about the source spectral and flux state in that time frame. The highest flux state in the soft and hard X-ray happened in Dec-2019, where the fluxes in the soft and hard X-ray bands are $\sim 9$ times that of their pre-CL value in 2011. In most cases we do not have signal to noise at energies $>5\kev$. However, to be consistent with the literature we quote the fluxes in the $2-10\kev$ band, which is the flux obtained by extrapolating the model to $10\kev$.

 The ratio between the X-ray and UV, which we refer to as $\alpha_{\rm OX}$, is calculated from the ratio of the monochromatic fluxes, i.e., $\alpha_{\rm OX}=-0.385\log(\rm F_{2\kev}/F_{2500\rm \angs})$ \citep{lusso2010}. This is an important diagnostic parameter to understand if the accretion disk and the X-ray emitting corona are physically connected. However, we note that we do not have UV fluxes exactly at $2500\angs$. We mostly use UVW2 ($1928\angs$) and UVM2 ($2246\angs$), and hence we extrapolated the fluxes obtained at these wavelengths to $2500\angs$ assuming a flat spectral slope. Our assumption of a flat is slope is valid, as we find from Table~\ref{Table:xray_obs} that in observations S23-S27 where we used UVM2 the flux densities are similar to those of S21-S21 and S28-S29 where we used UVW2. Table~\ref{Table:xray_obs} lists the $\alpha_{\rm OX}$ values at different epochs of the changing look state. Fig~\ref{fig:xray_uv_alpha_ox} captures the evolution of $0.3-2\kev$, $2-10\kev$ and $2500\angs$ fluxes, along with $\alpha_{\rm OX}$ over a period of $\sim 1400$ days post-flare. 

\subsection{\xmm{} EPIC-pn and Optical monitor}

We have analyzed the 2011 \xmm{} \citep{2001A&A...365L...1J} archival observation of the source 1ES~1927+654. The observation was taken during the pre-flare state of the source (See Table~\ref{Table:obs} for details). We used the latest XMM–Newton Science Analysis System (SAS v19.0.0) to process the Observation Data Files (ODFs) from all observations.  We preferred EPIC-pn \citep{2001A&A...365L..18S} over MOS \citep{2001A&A...365L..27T} due to its better signal to noise ratio. The EVSELECT task was used to  select  the single and double events for the pn detector.  We created light curves from the event files for each observations to account for the high background flaring used a rate cutoff of $<0.4$ counts/s. We also checked the pile up using SAS task {\tt epatplot}, and found that none of the obervations had any significant pile-up. Source and background photons were extracted from a circular region of 40 arc sec centred on the source and away from the source but on the same CCD respectively. The response matrices were generated using the SAS tasks {\tt arfgen} and {\tt rmfgen}. The spectra were grouped using the command {\tt specgroup} with a minimum count of 20 in each energy bin.

The Optical Monitor [OM]\citep{2001A&A...365L..36M} also collected data during this observations. We did not consider the V and B band due to possible host-galaxy contamination and obtained the count rates in four active ﬁlters (U, UVW1, UVM2, and UVW2) by specifying the R.A. and dec of the source in the source list ﬁle obtained by the {\sc omichain} task. In Table \ref{Table:xray_obs} we list the UVW2 mono-chromatic flux densities.


\subsubsection{\xmm{} EPIC-pn spectral analysis}
\label{subsubsection:XMM-analysis}

The 2011 XMM-Newton observation (pre-flare) was previously studied by \cite{Gallo2013}.  The authors found that the source is unobscured ($\rm \le N_{H} \sim 10^{20} \cmsqi$) and the X-ray spectrum could be well described by a power-law ($\Gamma = 2.27\pm0.04$) and a single black body component ($\rm kT_{e}=170\pm5\ev$). We followed the similar approach and found best-fit parameter values similar to them ($\Gamma = 2.21\pm0.02$ and $\rm kT_{e}=200\pm10\ev$). See Table~\ref{Table:xray_obs}. 


\subsection{The TNG optical observation in 2011}

\subsubsection{Data reduction}
The pre-changing look optical spectrum of 1ES~1927+654 was taken on 2nd~June~2011 (PI: Panessa), using the
DOLORES (Device Optimized for the LOw RESolution) instrument at TNG (Telescopio Nazionale Galileo). Two different grisms have been employed: the low LR-B (R$=600$ or $460 \kms{}$ around H$ \beta $) and the high VHR-R ($R=2500$) resolution, with the same 0.7$''$ slit. The wavelength range is 3600-8100 $\angs$ for the LRB and 6200-7700 $\angs$ for the VHR-R. The exposure time was of 300 sec and 1500 sec, respectively.

The data were reduced using standard processing techniques using MIDAS and IRAF packages. The raw data were bias subtracted and corrected for pixel-to-pixel variations (flat-field). Object spectra were extracted and sky subtracted. Cosmic-rays were removed. Wavelength calibrations were carried out by comparison with exposures of Ar, and Ne+Hg+Kr lamps. Flux calibration was carried out by observations of the spectro-photometric standard star Feige34 during the same night, with the same instrumental set-up, and by correcting for atmospheric extinction.

\subsubsection{Stellar absorption correction and emission line measurements}
We detect significant host galaxy stellar absorption in the TNG spectrum. The line H$\beta$ is seen in absorption with a complex profile, formed by a dominant absorption part filled with an emission line. Therefore in order to properly measure the emission lines we modeled the stellar population of the host galaxy. We have applied pPXF to fit the stellar population and subtract it to obtain a spectrum containing pure emission lines. During the pPXF process all the known emission lines are masked to avoid interference with the stellar population determination. Following the host galaxy stellar absorption subtraction, we modeled the emission line features using Gaussian components. Several constrains were applied during the fitting process: the line center offsets and relative intensity of the doublets [OIII]$\lambda\lambda\, 4959, 5007$ and [NII]$\lambda\lambda 6548, 6584$ were imposed from the atomic values. The line widths of each doublet were also tied to have the same velocity width. The results are presented in Table \ref{Table:dolores_lines} and Fig \ref{fig:spec_GTC_and_TNG} left panel.

\subsubsection{Fitting the broad H$\alpha$ emission line} \label{subsubsec:broad_lines}
The source had been classically referred to as a true type-II AGN due to its lack of any broad emission line signature in the optical spectra, as well as lack of any line of sight obscuration \citep{Gallo2013}. However, interestingly this source exhibited a transient broad-line region during the CL event \citep{trak19}, indicating the presence of a BLR, which is probably not visible due to its low emissivity in normal times. Following this, we carefully searched for the signature/hint of any weak broad line in the 2011 pre-CL TNG spectrum, which could tell us that the BLR existed all through, but not bright enough to be seen. After correcting for the host galaxy stellar absorption, we found that most emission lines exhibit narrow profiles with FWHM$\le 500\kms$ (See Table \ref{Table:dolores_lines}). However, we found some broad positive residuals in the region around the H$\alpha$ region after we have accounted for the narrow emission lines. 

To further investigate the broad residuals, we used an additional broad Gaussian (FWHM and normalization free to vary) over and above the narrow Gaussian used for the H$\alpha$ profile. We detected statistically significant improvement in the fit, and detected a weak broad H$\alpha$ emission line profile with FWHM$=2621\pm700\kms$, and line flux of $(2.10\pm1.25)\times10^{-15}\funit$. See Table \ref{Table:dolores_lines}. A close-up for the most complex emission line fits, H$\beta$ and H$\alpha$, can be found in Fig.~\ref{fig:spec_LRB_zoom}.


\subsection{Gran Telescopio CANARIAS}

The optical spectra of 1ES~1927+654 were observed with the 10.4\,m Gran Telescopio CANARIAS (GTC) under Director's Discretionary Time program GTC2021-176 (P.I. J. Becerra). Two observation epochs were carried out in order to investigate possible evolution, the first observation was performed on 10 March 2021 and the second one on 4 May 2021, as reported in Table~\ref{Table:obs}. 

The instrumental setup for both observations used OSIRIS in spectroscopic long-slit mode, with the grism R1000B and slit width of 1$''$, which yields a spectral resolution of 625 at 5000\,\AA. Three exposures of 150\,s each were taken on the target, resulting on a total of 450\,s per observation. The observations were performed in parallactic angle (${\rm PA} \simeq 110^\circ$ and ${\rm PA}\simeq 125^\circ$ on March and May 2021. The calibration stars G191-B2B and Ross640 were observed using the same instrumental setup for the first and the second observation respectively. Standard calibration images for bias, flat field and calibration lamp were also taken during the same night.

The GTC data were reduced following the standard procedures for bias subtraction, flat-field correction and sky subtraction. The flux calibration was performed using a mean response estimated from both calibration stars. The resulting flux calibrated spectra are shown in Appendix A. The spectra taken during the two epochs are compatible within the noise, ruling out evolution in the optical band. Therefore, the following analysis is performed using the combined spectrum corrected for interstellar reddening using the extinction curve from \cite{1999PASP..111...63F}. A quantitative justification of combining the two epochs of GTC spectra is given in Appendix A. We also make an elaborate comparison of the TNG (2011) and GTC (2021) spectra in Appendix A.

\begin{figure*}
    \centering
    \hbox{
    \includegraphics[width=8.5cm, angle=0]{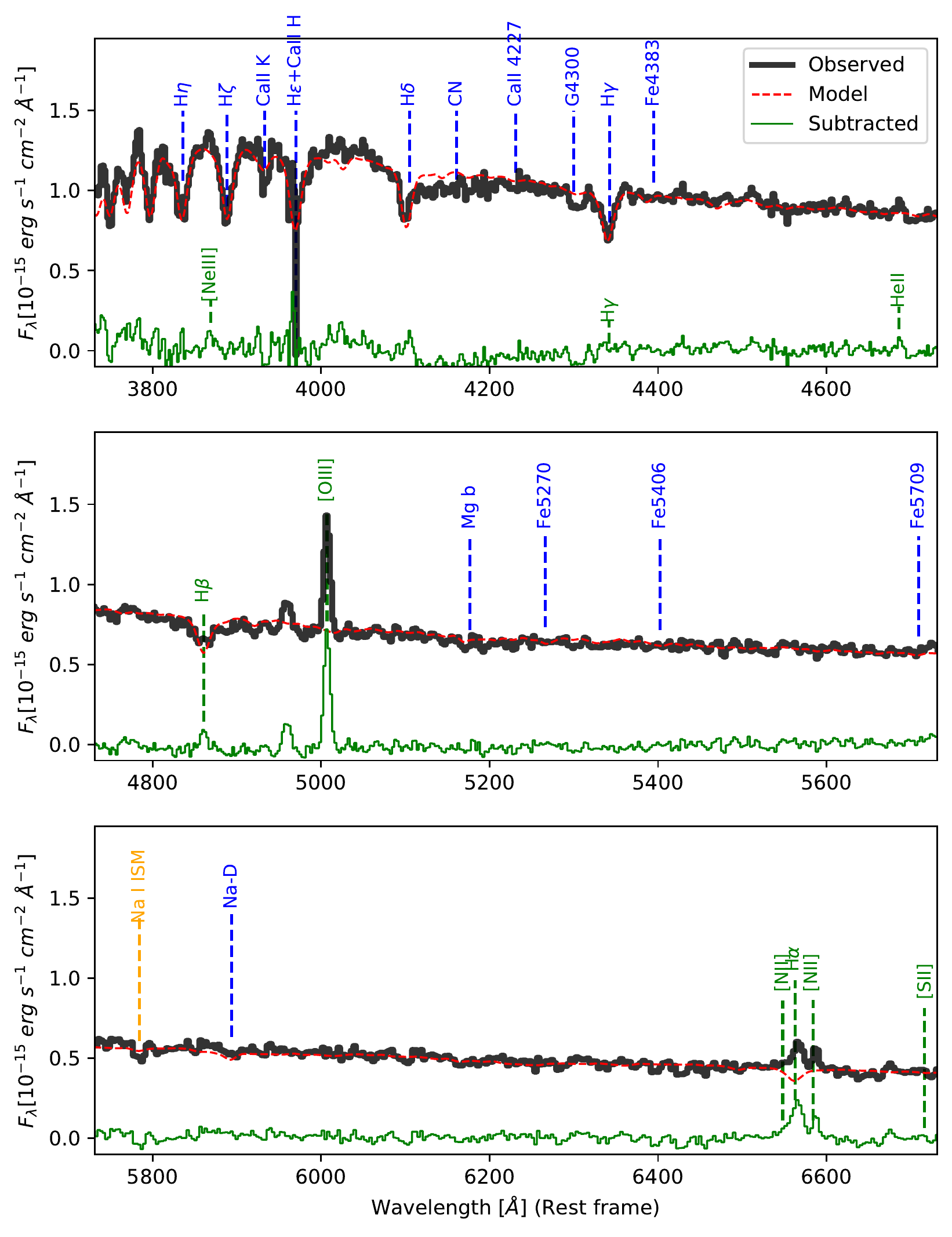}
    \includegraphics[width=8.5cm]{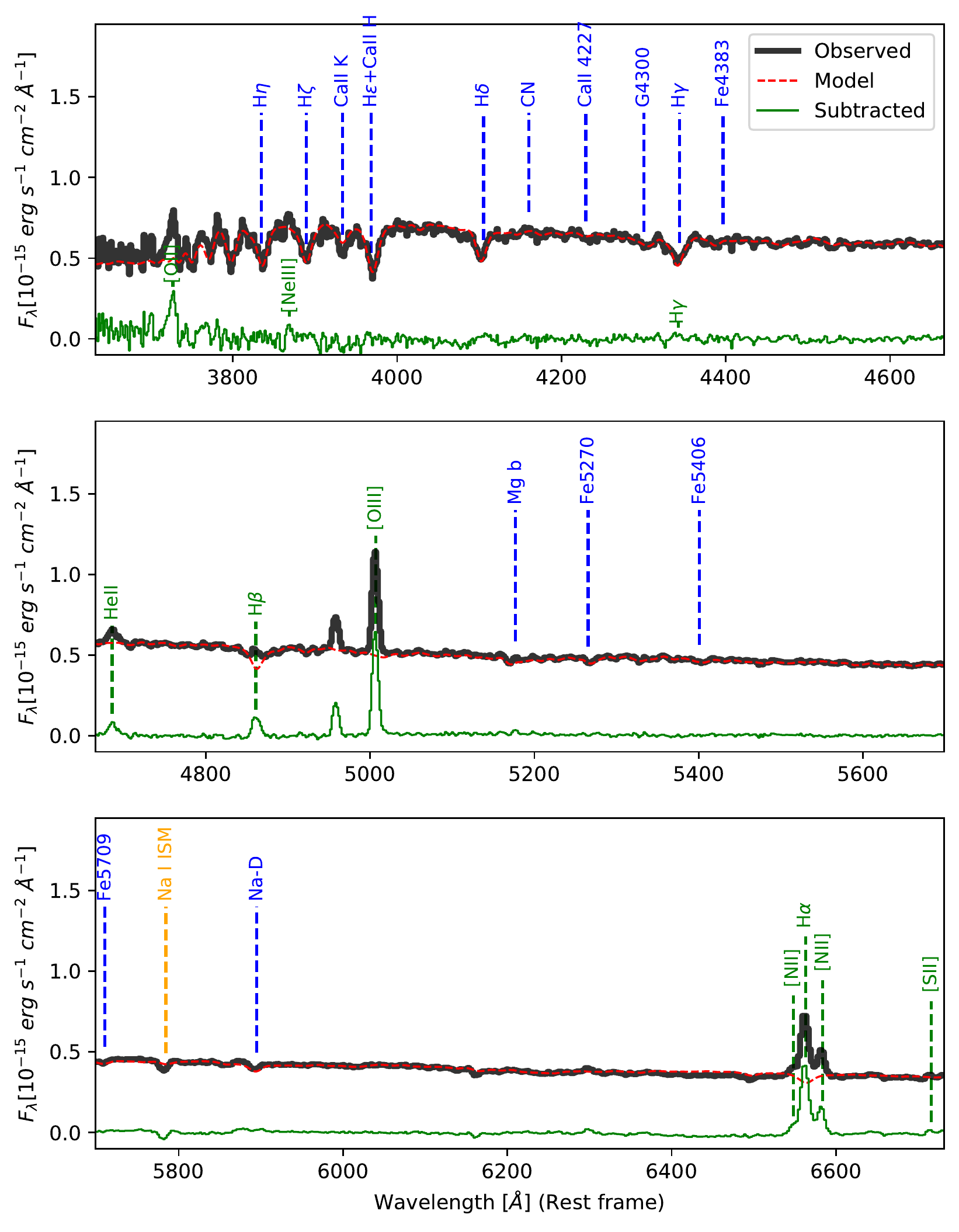}
    }
    \caption{{\it Left:} The pre-CL (2011, obtained with TNG) and {\it Right:} post-CL (2021, obtained with GTC) optical spectra of 1ES~1927+654. The black line represents the observed spectrum, the red dashed line denotes the stellar emission model and the green solid line shows the spectrum after stellar model subtraction. The most prominent absorption lines are labelled in blue, while the emission lines are labeled in green. Refer to Tables \ref{Table:dolores_lines} and \ref{tab:gtc_lines} for the emission line measurements. }
    \label{fig:spec_GTC_and_TNG}
\end{figure*}

The stellar population emission is present in the observed spectrum, in order to remove such contribution the penalized pixel fitting technique (pPXF) \citep{Cappellari04, Cappellari17} is used. We adopted the stellar population synthesis models MILES\footnote{Medium--resolution Isaac Newton Telescope Library of Empirical Spectra} \citep{Vazdekis10} with a range of metallicity from -0.2 to 0.5 and age from 30 Myr to 15 Gyr as template spectra. The combined observed spectrum together with the spectrum after stellar component subtraction can be found in Figure \ref{fig:spec_GTC_and_TNG} right panel.

After subtracting the stellar contribution, the emission lines detected in the optical spectra together with their basic characteristics are given in Table~\ref{tab:gtc_lines}. A zoom for the most complex emission line fits, H$\beta$ and H$\alpha$, can be found in Fig.~\ref{fig:spec_GTC_zoom}. We confirm that we do not detect any weak broad emission line in the H$\alpha$ complex in the GTC spectra. From the centers of the emission lines a redshift of z$=0.01878\pm0.00002$ is estimated. The errors are estimated by performing Monte-Carlo simulations of the observed spectrum with the uncertainty estimated from the root-mean-square (RMS) of the continuum.


\begin{figure*}
    \centering
    \includegraphics[width=15cm]{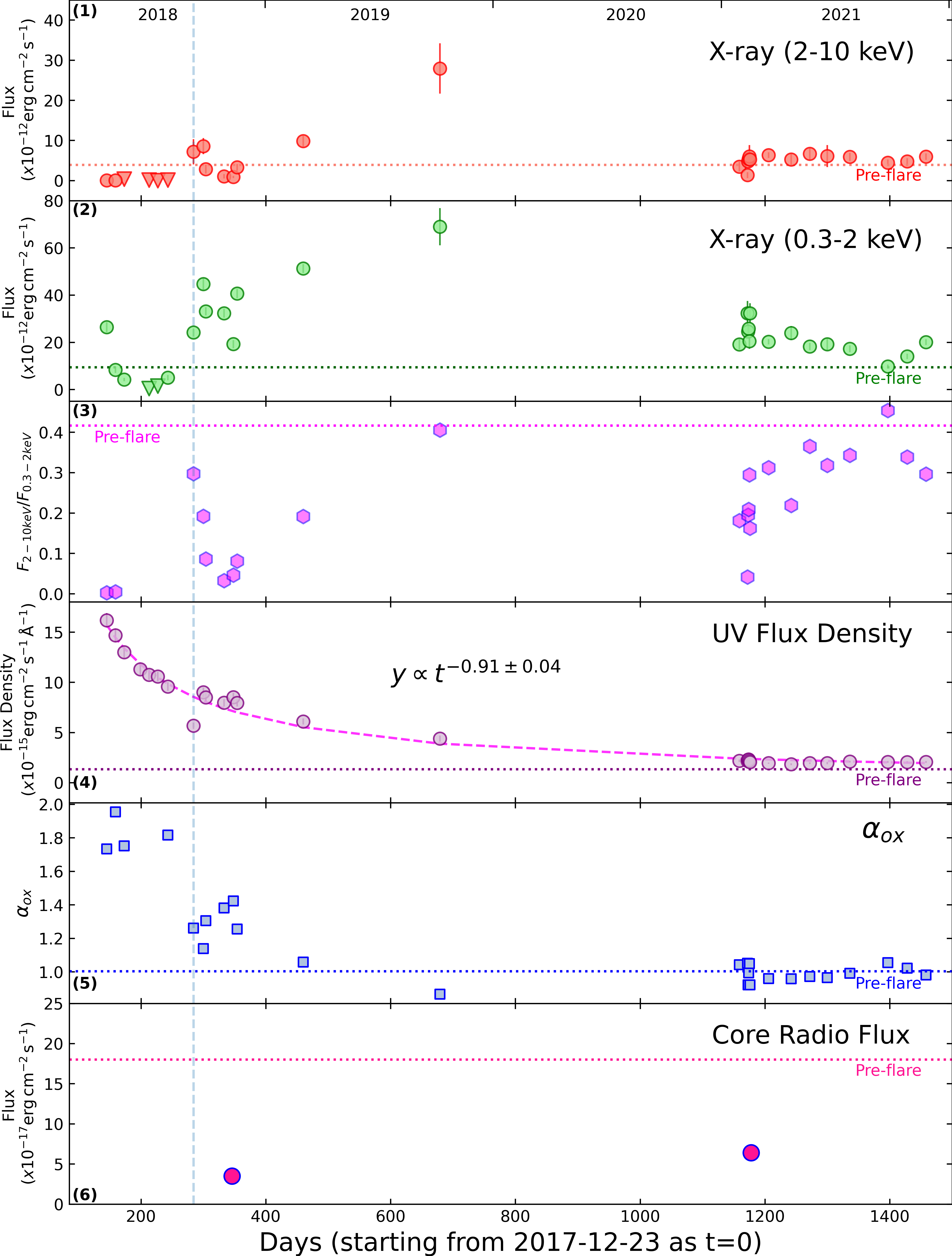}
    \caption{ The light curves of the X-ray and UV parameters of the central engine of the AGN 1ES1927+654, as observed by Swift (see Table \ref{Table:xray_obs} for details). The start date of the light curve is 2017-12-23 corresponding to the burst date reported by \cite{trak19}. The X-axis is in the units of days elapsed from the start date. The dotted horizontal lines in every panel refer to their pre-flare values (in 2011). The inverted triangles are the upper limits.  From the top to the bottom are panels: (1) The X-ray $2-10\kev$ flux (in units of $10^{-12}\funit$), (2) The X-ray $0.3-2\kev$ flux (in units of $10^{-12}\funit$), (3) The hardness ratio: $F_{2-10\kev}/F_{0.3-2\kev}$, (4) The UV (UVW2) flux density (in units of $10^{-15}\funita$), (5) the $\alpha_{\rm OX}$, and (6) the Core radio flux ($<1\pc$ spatial resolution){ NOTE: The vertical line corresponds to the the observation S8, where the X-ray corona jumps back (created) after being destroyed, and there is a dip in the UV flux by a factor of 2, and also the X-ray spectra becomes harder (panel 3). There could be a connection between the X-ray corona revival and the dimming of UV emission.} }
    \label{fig:xray_uv_alpha_ox}
\end{figure*}



\begin{table*}
    \begin{center}
        
    \caption{Pre-changing-look optical spectral features, observed in 2011 by TNG\label{Table:dolores_lines}}

    \begin{tabular}{lccc}
       \hline
       \hline
   
Line & wavelength & FWHM & Line Flux \\ 
     & ($\angs$) & ($\kms$) &  ($\funit{}$)\\
     &  &  &   ($ \times 10^{-15}$)\\
\hline

$\rm H\beta$ & $4952.1 \pm 0.7$ &  $555/310 \pm 135$  &  $1.25 \pm 0.37$ \\
$\rm [OIII]4959$ & $5052.45 \pm 0.11$ &  $521/261 \pm 14$  & $2.25 \pm 0.08$ \\
$\rm [OIII]5007$ & $5101.29 \pm 0.11$  &  $521/268 \pm 14$ & $6.81 \pm 0.24$ \\
$\rm [NII]6548$ & $6668.4 \pm 1.0$ &  $417/240 \pm 140$ & $0.51 \pm 0.18$ \\
$\rm H\alpha$\tablenotemark{\footnotesize A} & $6682.6 \pm 0.51$ &  $721/636 \pm 70$  & $4.09 \pm 0.43$\\
$\rm [NII]6584$ & $6704.4 \pm 1.0$ &  $417/242 \pm 140$ & $1.55 \pm 0.53$ \\
\hline\\
\end{tabular}

\hspace{1.5cm}{H$\alpha$ profile modelled by a broad and a narrow component.}\\

\begin{tabular}{lccc}
\hline 
$\rm [NII]6548$ & $6668.0 \pm 0.5$  &  $421/247 \pm 14$ & $0.43 \pm 0.07$  \\
$\rm H\alpha-n$\tablenotemark{\footnotesize B} & $6683.0 \pm 0.5$ &  $625/524 \pm 60$ & $3.13 \pm 0.49$  \\
$\rm H\alpha-b$\tablenotemark{\footnotesize B} & $6673.18 \pm 15$ & $2644/2621 \pm 700$ & $2.10 \pm 1.25$  \\
$\rm [NII]6584$ & $6704.49 \pm 0.69$ & $421/249 \pm 85.3$ & $1.29 \pm 0.22$  \\

        \hline 
        \hline
    \end{tabular}
    \end{center}

\tablenotetext{\footnotesize A}{With a single Gaussian line fit to the H$\alpha$ complex we found that the line is slightly broad.}
  
\tablenotetext{\footnotesize B}{The H$\alpha$ emission line was modelled using two Gaussian components for H$\alpha$: narrow H$\rm \alpha-n$ and broad H$\rm \alpha-b$. In this case, the line width of [NII]$6548, 6584$ were fixed to the value obtained with a single component.}  

\tablecomments{Measurements of the spectral features in the pre-changing-look optical spectrum, observed in 2011 by TNG/LRB. The spectrum has been corrected for Galactic extinction. The host galaxy stellar emission has been subtracted.  The two line width values correspond to the measured/deconvolved by the instrumental profile. The complex H$\alpha+[NII]$ has been fitted in two modes: first using a single component for each spectral feature, and second using two components for H$\alpha$, quoted below the horizontal line. In the second mode the line width for [NII]$6548, 6584$ were kept fixed to the value obtained in the first case.}    
    
\end{table*}

\begin{table*}
    \begin{center}
    \caption{The post-changing-look optical spectral features observed by GTC/OSIRIS in 2021  \label{tab:gtc_lines}}
    \begin{tabular}{lcccc}
       \hline
       \hline
       Line ID  & Center & FWHM & Line Flux & Comment$\rm ^B$  \\ 
     &   (\AA)  & ($\rm km\,s^{-1}$) & ($\funit{}$)  \\
     &    &  & $\times 10^{-15}$  \\
     
        \hline
$\rm [OII]3727$ & $3796.8 \pm 0.6$ & $710/285 \pm 100$ &  $2.34 \pm 0.4$ & new line  \\
$\rm HeII4686$ & $4775.1 \pm 0.4$  & $795/603 \pm 80$  &  $0.99 \pm 0.11$ & new line \\
$\rm H\beta$ & $4953.3 \pm 0.4$ & $660/435 \pm 30$ &  $1.43 \pm 0.08$  &same \\
$\rm [OIII]4959$ & $5052.09 \pm 0.03$ & $505/125 \pm 4$ & $1.95 \pm 0.02$ &lower ($-15\%$) \\
$\rm [OIII]5007$ & $5100.92 \pm 0.03$  & $505/143 \pm 4$ &  $5.91 \pm 0.06$ & lower ($-13\%$) \\
$\rm [OI]6300$ & $6417.56 \pm 0.82$ & $477/281 \pm 86$ & $0.29 \pm 0.07$ & new line\\
$\rm [NII]6548$ & $6670.00 \pm 0.19$ & $500/335 \pm 22$ &  $0.66 \pm 0.04$ & higher ($+22\%$)  \\
$\rm H\alpha$ & $6686.22 \pm 0.08$  & $544/400 \pm 9$ &  $5.64 \pm 0.13$  &higher ($+18\%$) \\
$\rm [NII]6584$ & $6706.06 \pm 0.20$ & $500/352 \pm 22$ &  $2.00 \pm 0.11$  &higher ($+22\%$) \\
        \hline 
        \hline
\end{tabular}
    
\end{center}

\tablecomments{ The post-changing-look optical spectral features measured from the combined spectrum after stellar template subtraction.  The spectrum has been corrected for Galactic extinction. 
The center and width of the lines [OIII]$ 4959,5007\angs \angs$ and [N$\sim$II]$ 6548,6584\angs\angs$  are tied. The line flux ratios [OIII]$5007/4959$ and [N~II]~$6584/6548$ are kept fixed to the theoretical value 3. \\
{$\rm ^B$ Listing the instances when the lines in the post-CL (2021) spectra have lower, higher or similar fluxes compared to the pre-CL spectra (2011) reported in Table \ref{Table:dolores_lines}}. }
The two line width values correspond to  the measured/deconvolved by the instrumental line profile.

\end{table*}


\subsection{VLA}
The source 1ES~1927+654 has been observed twice by the Very Large Array, at C-band (BC configuration) and X-band (AB configuration). Both data sets were reduced using the Common Astronomy Software Applications (\texttt{CASA}) package version 5.3.0 \citep{casa}.  The C-band observation was taken on 1992 Jan 31 and consists of one 3.5 minute scan. Standard calibration was applied with source 1959+650 serving as the initial amplitude and phase calibrator and 3C~48 as flux calibrator. Imaging deconvolution was conducted using the task \texttt{clean} as implemented in \texttt{CASA}, with a Briggs weighting and robust parameter of 0.5. One round each of phase-only and amplitude and phase self-calibration was applied. The resulting image has an RMS of 1.25$\times10^{-4}$~Jy/beam and the synthesized beam is 4.51$''\times$1.48$''$.  The radio emission at C-band is unresolved (i.e., a point source); a Gaussian fit to the central peak gives a flux of 16.4$\pm$0.2~mJy at a central frequency of 4.86 GHz, in agreement with the value of 16$\pm$2~mJy previously published in \cite{perlman1996}.

The X-band observation was taken on 1998 June 06 and consists of two 2-minute scans. The same procedure was followed as for C-band, with source 1800+784 serving as initial phase calibrator. The final image has an RMS of 9.26$\times10^{-5}$~Jy/beam and synthesized beam of 0.47$''\times$0.22$''$. The source is not resolved, and the limits on the source size from the Gaussian fit tool in \texttt{CASA} is less than 88$\times$30 mas, or 34$\times$12 pc. The peak flux of the point source is 9.12$\pm$0.11 mJy at a central frequency of 8.46 GHz.

\subsection{VLBA}
\label{methods:vlba}
1ES 1927+6547 was observed by the Very Long Baseline Array, concurrent with the \emph{Swift} campaign, on 15 March 2021 under Director's Discretionary Time proposal 21A-403 (PI: E. Meyer). A standard dual-polarization 6 cm frequency setup was used, with central channel frequencies of 4868 MHz, 4900 MHz, 4932 MHz, 4964 MHz, 4996 MHz, 5028 MHz, 5060 MHz and 5092 MHz and a total bandwidth of 32 MHz. As the source was expected to be too faint for self-calibration, we used a relatively fast-switching cadence between the target and a bright calibrator source, J1933+6540, 1.2$^\circ$ distant. Target scans were 190 seconds with 30 seconds on the calibrator. We observed J2005+7752 and J1740+5211 as amplitude check sources. The entire observation was 3.5 hr resulting in acceptable UV coverage for imaging.

\begin{figure}
    \centering
    \includegraphics[width=\linewidth]{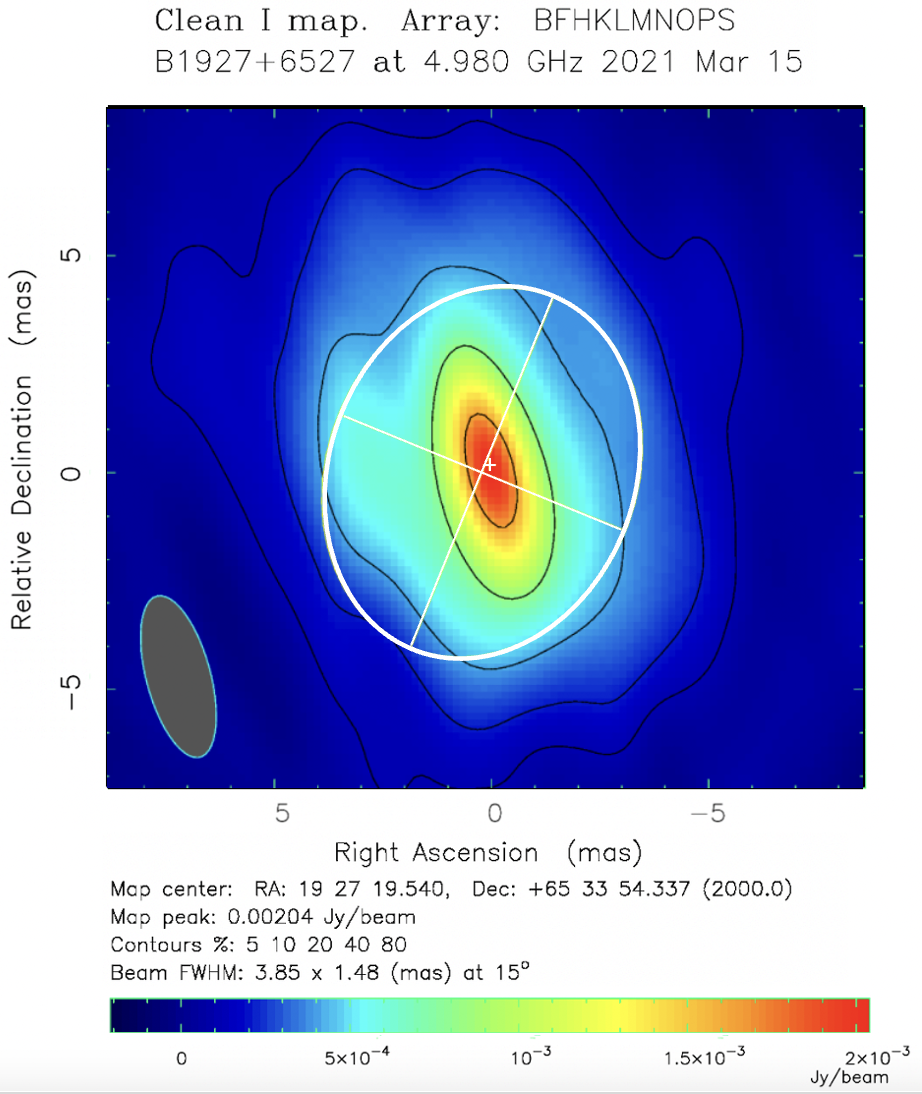}
    \caption{Image corresponding to the 2021 VLBA observation of 1ES~1927+654, at 4.9 GHz. Extended diffuse emission, in addition to a bright core, is evident. The best-fit point-source+disk model has been illustrated with the point source shown as a small white cross while the disk as the white oval with corresponding major and minor axes. The contours have been plotted as 5, 10, 20, 40 and 80\% of the peak flux. The extended disk model cannot be convolved with the synthesized beam inside \texttt{DIFMAP} and hence this \texttt{CLEAN} map is only the core+residual.}
    \label{fig:vlba}
\end{figure}

The data were checked for radio frequency interference and then calibrated using the rPicard pipeline \citep{janssen19} installed in Common Astronomy Software Applications (CASA) version 5.6.0. Initial imaging deconvolution was accomplished in \texttt{DIFMAP} \citep{shep94} with a map size of 1024 pixels at 0.16 mas/pixel. The restored image shown in Figure~\ref{fig:vlba} used natural weighting and has an RMS sensitivity of 0.5 $\mu$Jy/beam and synthesized beam (resolution) of 3.85 $\times$ 1.48 mas. The imaging shows a central peak of approximately 2 mJy/beam. The total flux is 5.5$\pm$0.5~mJy, suggesting a resolved component, and extended emission around the point source is also apparent in the residual imaging.  To determine better the true source structure we used a custom version of \texttt{DIFMAP} \texttt{modelfit} (\texttt{ngDIFMAP}, \citealt{royc21}) to fit the visibilities of the calibrated data with several alternative models including a single unresolved point source and either adding a Gaussian or a uniform disk model to the same. The best model (as determined by the reduced chi-squared) is a point source of 1.4~mJy centered on a uniformly bright disk of 4.1~mJy and size 4.4$\times$3.5 mas, corresponding to 1.8 by 1.4 pc (see Table~\ref{Table:radio}). We used Monte Carlo simulations \citep[e.g.,][]{briggs95,chael18,royc21}, to verify that the extended emission around the point source (shown as a small cross in Figure \ref{fig:vlba}) is intrinsic to the source and is not an artifact of interferometric errors or gaps in $u-v$ coverage.

\subsection{EVN}
\label{methods:evn}
1ES 1927+6547 has been observed by the European VLBI Network (EVN) under project EG079 in March 2013 and October 2014 at 1.5 and 5 GHz respectively, and under project RSY07 in December 2018 at 5 GHz. These epochs will be referred to by year (2013, 2014, and 2018) in the rest of the paper. We note that the first two observations pre-date the CL event while the 2018 observation was about 1 year after the optical/UV peak. We used the publicly available pipeline-reduced \texttt{uvfits} files\footnote{archive.jive.nl/scripts/avo/fitsfinder.php} for our analysis. The source was observed for $\sim$3-4 hours for each of the EVN observations, similar to the VLBA 2021 observation, implying expected sensitivities\footnote{http://old.evlbi.org/cgi-bin/EVNcalc} $\sim10-20\mu$Jy/beam for all the VLBI data. However, $\sim$3-4 antennas dropped out for each of the EVN observations, which worsened $u-v$ coverage, and therefore sensitivity.

We used \texttt{DIFMAP} for initial imaging deconvolution with a map size of 1024 pixels at 0.1 mas/pixel. Using the same modified \texttt{DIFMAP} fitting methods as for the 2021 VLBA observation, we again evaluated alternative models for the radio emission. The best-fit model in both 2013 (1.5 GHz) and 2014 (5 GHz) is an unresolved point source, while the post-flare 2018 epoch is better described by a point source atop a uniformly bright disk, as in the 2021 VLBA observation. The best-fit model fluxes, the RMS sensitivity and the resolution (synthesized beam size) have been tabulated in Table \ref{Table:radio} along with all other radio observations described here. As noted, the fit results from 2018 and 2021 are similar. Because the EVN data suffered from poorer $u-v$ coverage due to several unusable antennas, we have carefully evaluated the uncertainties in the model parameters through Monte Carlo simulations of the visibilities \citep{royc21}. For the EVN observations we further assume a minimum $10\%$ error on the flux due to uncertainties in the flux calibration.

 \begin{table}[]
    \centering
    \caption{The optical diagnostic emission line ratios.\label{tab:BPTratios}}
    \begin{tabular}{lcc}
    \hline \hline
        LineRatio & 2011 & 2021  \\
        \hline
        H$\alpha$/H$\beta$ &  3.3$\pm$1.15 &  3.94$\pm$0.3 \\          
        {[}OII{]}/H$\beta$ &  ... & 1.6$\pm$0.3   \\
        {[}OIII{]}/H$\beta$ &  5.47$\pm$1.8 &  4.14$\pm$0.2 \\
        {[}OI{]}/H$\alpha$ &  ... & 0.051$\pm$0.012   \\
        {[}NII{]}/H$\alpha$ &  0.38$\pm$0.13/0.25$\pm$0.08$^a$ &  0.36$\pm$0.02 \\
        \hline
    \end{tabular}
    {Line ratios are computed after stellar contribution has been subtracted. Note that in either 2011 or 2021, the H$\beta$ emission line was not detectable unless we subtracted the underlying stellar absorption.
    $^a$ Line fluxes were computed assuming a broad component for H$\alpha$ as described in previous sections.}
    
\end{table}

\begin{table*}
\centering
  \caption{Details of the radio observations, with corresponding flux density measurements.  In cases where only an unresolved core is observed, the total flux density equals the core flux density.  For cases where we detect resolved extended emission, the central point source (PS) flux is noted alongside the extended flux density and the semi-major and semi-minor axes of a best-fitting uniform disk model. Note that ``flux$"$ in the table headers refers to flux density. The brightness temperature $T_B$ has been calculated for the point source at 5 GHz. }\label{Table:radio}
  \begin{tabular}{llllllllll} \hline\hline 

Obs. & Freq. & Date & Total flux & Central PS flux &  Extended flux & Disk Dimensions &  RMS & Resolution & $T_B$ \\
& (GHz) & {\footnotesize (MM/YY)} & {\footnotesize (mJy/beam)} & (mJy) &  (mJy) & (mas$\times$mas) & (Jy/beam) & (mas$\times$mas) & ($\times 10^6$ K) \\
VLA & 8.46 & 06/98 & 9.0$\pm$0.1 & --- &  --- & --- & $1.3\times10^{-4}$ & 470$\times$220 & ---\\
VLA & 4.86 & 01/92 & 16.4$\pm$0.2 & --- &  ---  & --- & $9.3\times10^{-5}$ & 4510$\times$1480 & ---\\
EVN & 4.99 & 03/14 & 4.1$\pm$0.4 & --- &  ---  & --- & 3.2$\times10^{-5}$ & 2.47$\times$1.18 & $15.0\pm1.5$ \\
EVN & 4.99 & 12/18 & 8.4$\pm$0.8 & 0.8$\pm$0.1 &  7.6$\pm$0.7 & 4.9$\pm$0.9$\times$4.6$\pm$0.5 & 3.6$\times10^{-5}$ & 6.01$\times$4.95 & $0.50\pm0.05$ \\
VLBA & 4.98 & 03/21 & 5.5$\pm$0.5 & 1.4$\pm$0.1 &  4.1$\pm$0.4 & 4.4$\pm$0.1$\times$3.5$\pm$0.1 & 5.2$\times10^{-5}$ & 3.85$\times$1.48 & $5.7\pm0.6$ \\
EVN & 1.48 & 08/13 & 18.9$\pm$0.2 & --- &  --- & ---  & $2.5\times10^{-4}$ & 28.2$\times$11.7 & --- \\

\hline 
\end{tabular} 
\end{table*}


\section{Results}\label{sec:results}
In this section we present the observational results from our multi-wavelength campaign of the CL-AGN 1ES1927+654. We have used X-ray and UV observations from \swift{} and \xmm{}, radio observations from {\it VLA}, {\it EVN} and {\it VLBA}, and optical observations from the {\it TNG} and {\it GTC}.


\subsection{The X-ray and UV spectra and light curves}

\subsubsection{The UV light curve}

The UV flux density starts off from a high flux state in May 2018 and drops montonically until Feb 2021 (we do not have any observations from 2020), after which it plateaus at a value of $\sim (2.06\pm0.11) \times 10^{-15}\funita{}$, its current post-flare quiescent state. Table \ref{Table:xray_obs} shows that this value is still slightly larger than the pre-CL value of $1.34\pm0.03 \times10^{-15}\funita{}$. From Figure \ref{fig:xray_uv_alpha_ox} panel 4 we find that the UV light curve drops at a rate $\propto t^{-0.91\pm 0.04}$, which is a shallower slope than predicted by a TDE event, which is typically $\propto t^{-5/3}$ \citep{van2021}. 

\begin{figure*}[h!]
    \centering
    \hbox{
    \includegraphics[width=9.2cm]{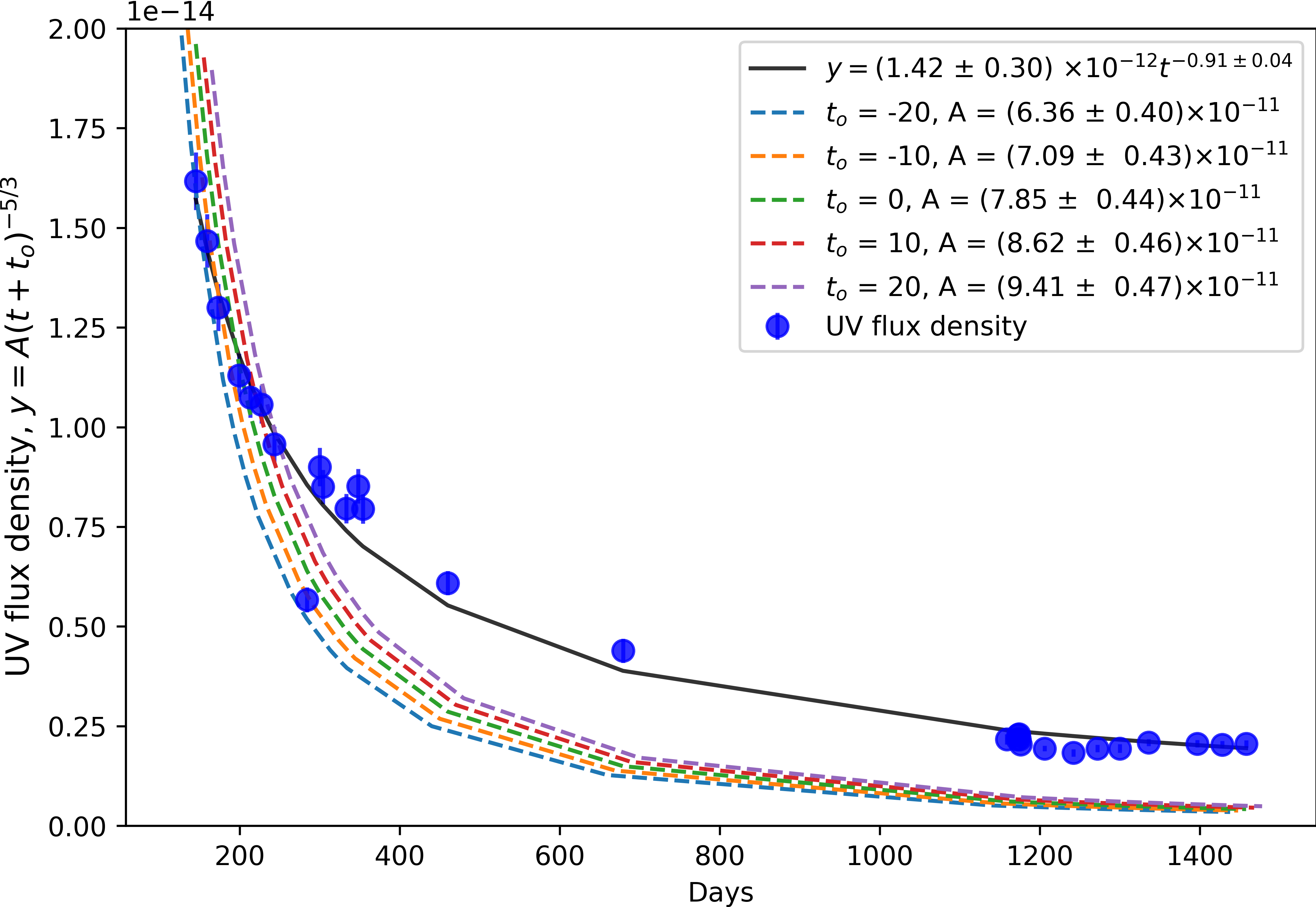}
    \includegraphics[width=9.2cm]{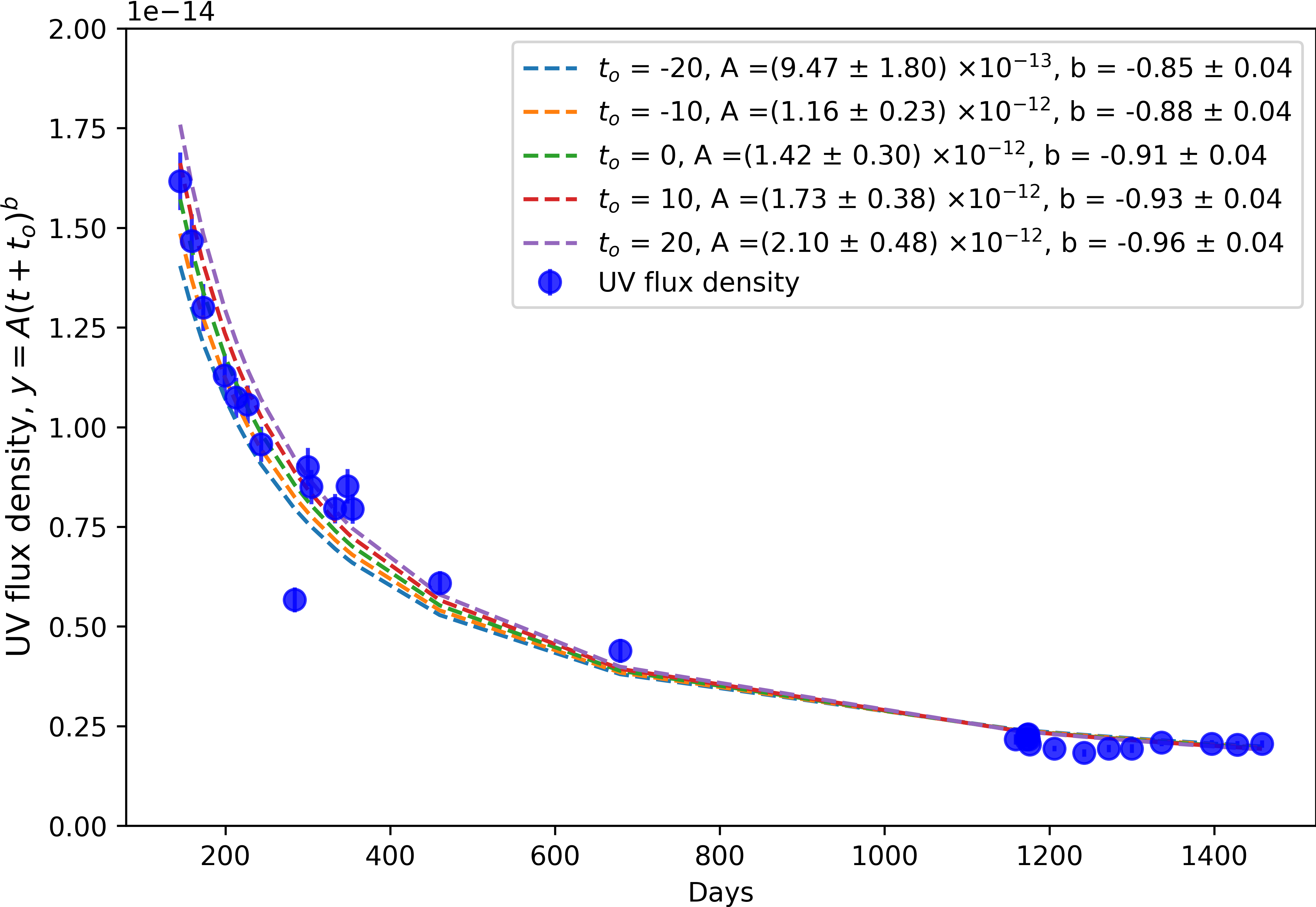}
    }
    \caption{{\it Left:}  The black curve represents the best fit to the UV flux density, with the equation $y=A \times (t+t_0)^{\rm b}$, where the starting time $t_0= 0$ is assumed to be 23rd Dec 2017 following \cite{trak19}. The blue points are the measured UV flux density (see Table \ref{Table:xray_obs}). The best fit values are $A= (1.42\pm0.30) \times 10^{-12}$, and $\rm b=-0.91\pm0.04$, for $t_0=0$. The colored dotted curves are the fits obtained when we froze $\rm b=-5/3$ following the predictions of a tidal disruption event (TDE). We allowed the normalization $A$ to vary, however, none of them gave a good description of the observed data. Due to the uncertainty in accurately knowing the day of the flare, we have fitted the data with different values of $t_0=-20,\, -10, \,0,\, 10, \, 20$, corresponding to 10 and 20 days around 23rd Dec 2017 (different colors). The fitted normalization $A$ in each case are reported in the figure. {\it Right:} Fitting the UV light curve with the equation $y=A \times (t+t_0)^{\rm b}$. In this case both $A$ and $\rm b$ are kept free. The different fitted curves correspond to the different start times (as in the left panel), due to the uncertainty in the start date. The best fit $A$ and the corresponding $\rm b$ values are reported in the figure.  }
    \label{fig:UV_lightcurve_fit}
\end{figure*}

In order to robustly test if the measured slope is indeed shallower, we did the following tests. We fitted a simple exponential function $y=A\times (t-t_0)^{b}$ to the UV light curve, and used least squares minimization technique to obtain the best fit, using the Python function \texttt{scipy.optimize.curve\_fit}. This function returns the best fit parameters and one standard deviation error on the parameters. For $t_0=0$ (i.e, a start date on 23rd Dec 2017), this results in a best fit slope of $b=-0.91\pm0.04$ and a normalization of $A= (1.42\pm0.30) \times 10^{-12}$ (See Figure \ref{fig:UV_lightcurve_fit}). As a next step, we froze the exponent value to that expected for a TDE (that is $b=-5/3$), and estimated the best fit by allowing the normalization A to vary, and for the different cases of the start date $t_0$, given that there is some doubt about which day the flare happened. We found that in all cases it did not describe the observed light curve at all (See Fig \ref{fig:UV_lightcurve_fit} left panel). The black solid curve on the same figure is the best fit with $b=- 0.91\pm 0.04$. Figure \ref{fig:UV_lightcurve_fit} right panel shows the different fitted curve for different start dates, where both the slope $b$ and normalization $A$ are left free to vary. Even with a spread of start date of $\sim 20$ days around the time we assumed (23rd Dec 2017), the best fit slopes are still consistent with $b\sim -0.9$.


\begin{figure}[h!]
    \includegraphics[width=8cm]{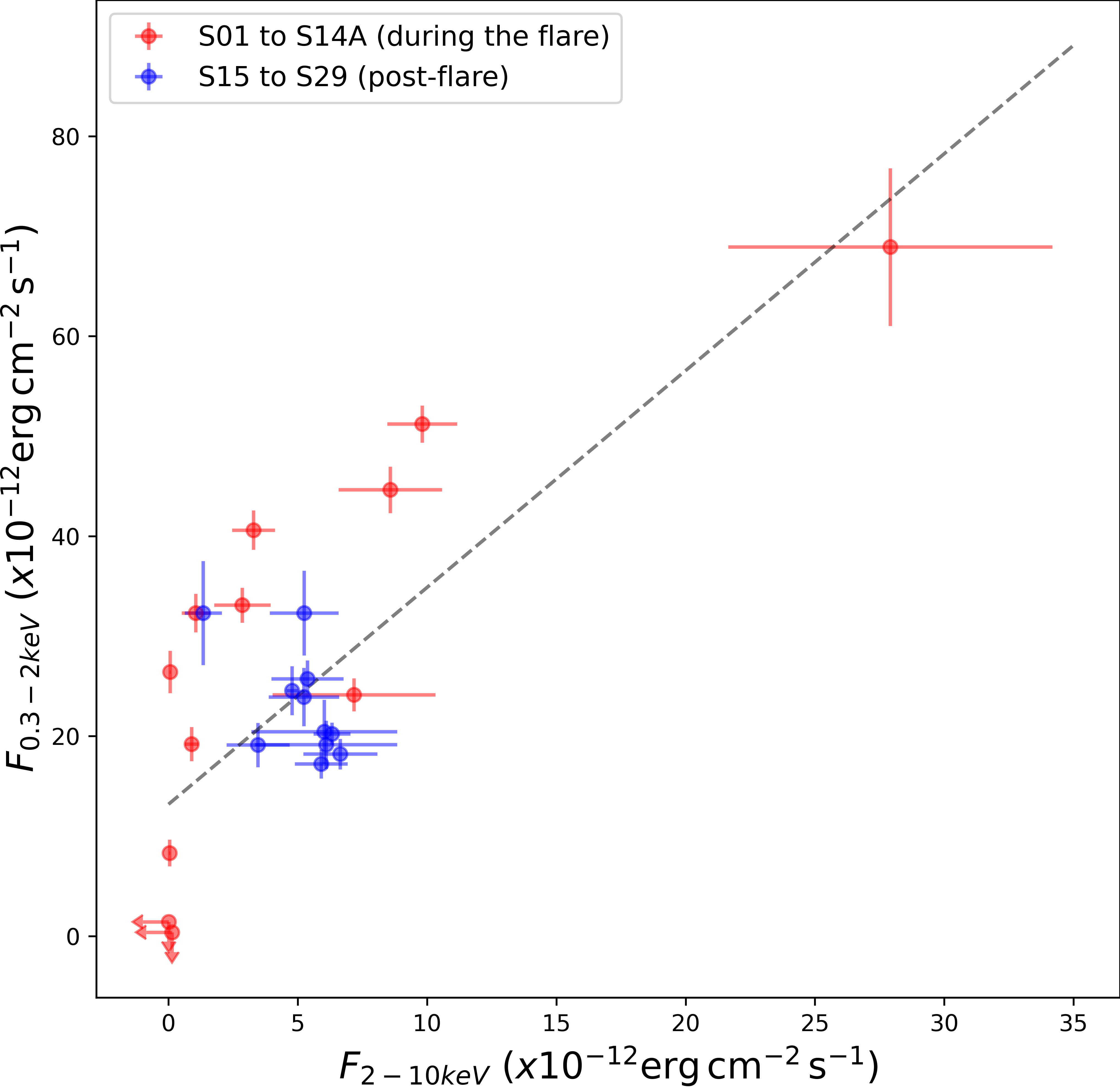}
    
    \caption{ The relation between the soft X-ray ($0.3-2\kev$) and hard X-ray ($2-10\kev$) fluxes at different epochs during and after the X-ray flaring period. The red circles denote the X-ray flaring period covering 2018-2019 (observations S01-S14A), while the blue circles denote the post-flare period covering Feb 2021- Oct 2021 (observations S15-S29). The lowest state in X-rays are denoted by upper limits in both the axes in the lower left corner of the figure. It is clear that the soft and the hard X-rays do not show any significant correlation (Spearman rank correlation coefficient=0.48, with a probability of rejecting the null hypothesis=98\%.}
    \label{fig:soft_hard_correlation}
\end{figure}

\begin{figure}[h!]
\includegraphics[width=9cm]{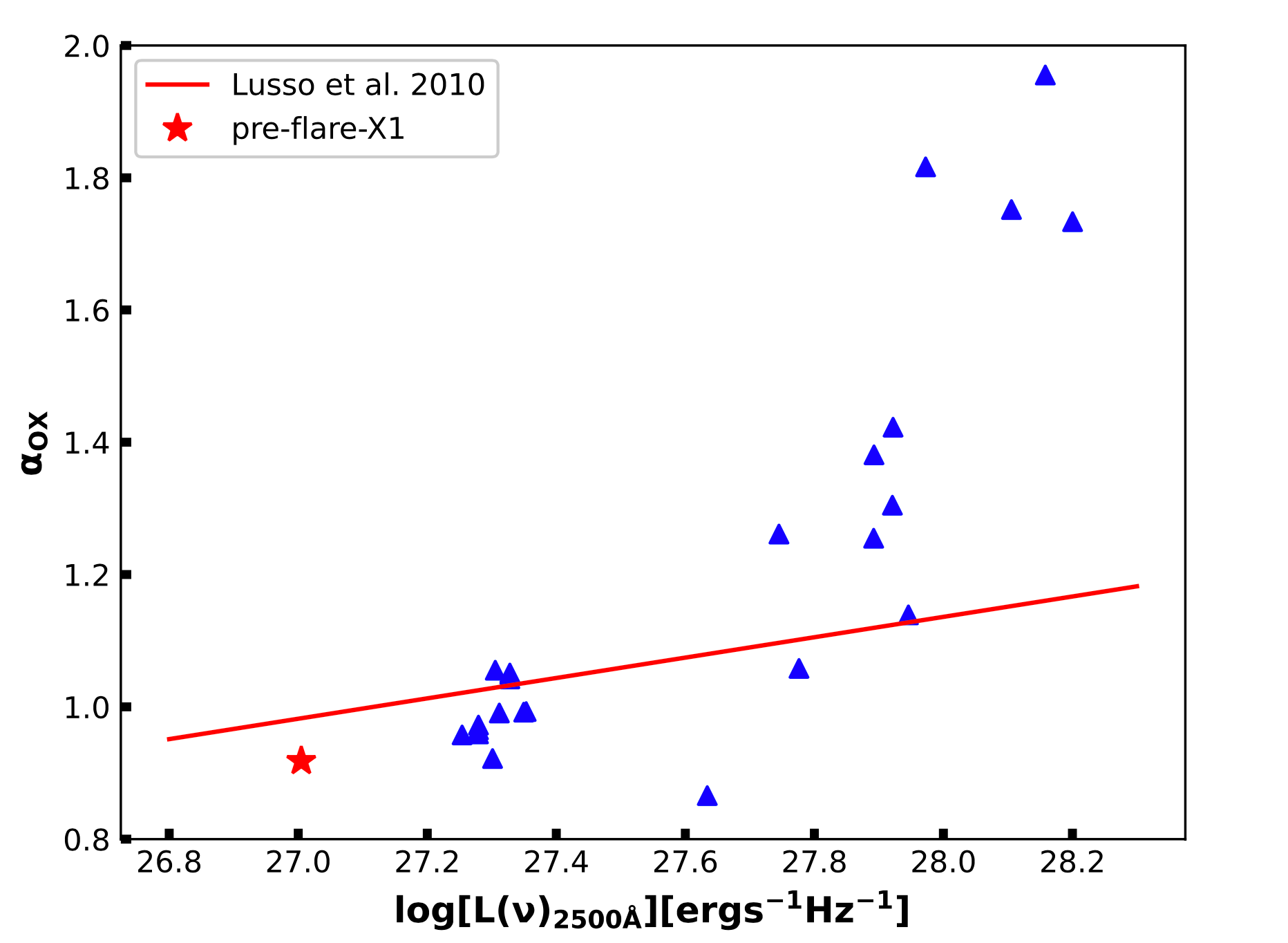}
    \caption{ The $\alpha_{\rm OX}$ vs $L_{2500\angs}$ for the different \swift{} observations of the source 1ES~1927+654 as reported in Table \ref{Table:xray_obs}. The red line denotes the best fit correlation obtained from \cite{lusso2010}, which denotes the standard AGN disk-corona relation. The red star denotes the 2011 pre-CL state of 1ES~1927+654. }
    \label{fig:alphaOX}
\end{figure}

\subsubsection{The X-ray light curve}

The X-rays however, behave in completely different manner, as also reported in the previous studies \citep{Ricci2020,Ricci2021}, where the flux drops to a minimum in June-Aug 2018 ($\sim 200$ days after the start of the flare) and then ramps up from Oct 2018 until it reaches its highest state in Nov 2019 (about 10 times its pre-CL value), and then falls back to its pre-CL state. The hard X-rays and soft X-rays do not show any correlated variability. See Figure \ref{fig:soft_hard_correlation}. We note that in the current post-flare state (2021) there is still some variability ($\sim$ factor of 2) in both the soft and hard X-ray emission. For example, the soft X-ray emission dropped exactly to its pre-CL value as late as Oct-2021, that is after $\sim 1400$ days from the start of the flare (See Figure \ref{fig:xray_uv_alpha_ox} panel 2). The $2-10\kev$ flux has reached its pre-CL state earlier.  From Table \ref{Table:xray_obs} we find that for all the X-ray observations, the soft X-ray ($0.3-2\kev$) flux dominates the overall X-ray luminosity. We also note that the spectra became very soft during the lowest flux states in 2018 (observations S01 to S07), consistent with what has been reported by previous studies \citep{Ricci2020,Ricci2021}. 

Figure \ref{fig:xray_uv_alpha_ox} panel 3 shows the hardness ratio (HR=$F_{2-10\kev}/F_{0.3-2\kev}$) light curve during and post-CL event. It is interesting to note that the HR does not follow the trend of the soft or hard X-ray fluxes. The HR reached its pre-CL state in Dec 2019, but again dropped to its minimal value in Feb 2021. The HR gradually revived to normal pre-CL state again in Oct 2021. These point to fact that the coronal and the soft X-ray emission are still in a phase of building up. 

Another interesting note is that the UV flux drops down by a factor of 2 (from its normal falloff) during the observation S8, coinciding exactly with the revival of the X-ray coronal emission after the violent event (plotted as a vertical dotted line in Fig \ref{fig:xray_uv_alpha_ox}). We find that the spectra becomes hard at that point (panel 3) for the first time after days of non-detection of hard X-rays. In Table \ref{Table:xray_obs} we find that we could obtain only upper limits in the $2-10\kev$ X-ray flux from S03 to S07 observations, and the hard X-ray emission revived in S8.

\subsubsection{The relation between $\alpha_{\rm OX}$ and $L_{\rm 2500\angs}$}

The universal relation between AGN accretion disk and corona is well described by the strong correlation found between $\alpha_{\rm OX}$ vs $L_{\rm 2500\angs}$, across large ranges of black hole mass, accretion rates and redshift \citep{lusso2010}. Fig \ref{fig:alphaOX} shows the $\alpha_{\rm OX}$ vs $L_{\rm 2500\angs}$ for all the \swift{}  observations reported in Table \ref{Table:xray_obs}. The red star in the lower left corner of the figure represents the pre-CL value, and interestingly it is slightly below the \cite{lusso2010} correlation (red line) indicating a dominant X-ray emission, relative to UV. The other data points obtained during the CL event are scattered all over the phase space indicating that the disk corona relation was not valid during this violent event. The recent observation data points (S25-S29) are clustered in the lower left corner (near the red star) of Fig \ref{fig:alphaOX} left panel, indicating that the disk-corona relation is gradually being established.




 \subsection{The optical spectra}
 
 \subsubsection{Comparing the pre and post changing-look spectra}

In order to check carefully any changes that may have been detected in the pre-CL and post-CL optical spectra, we carried out a uniform analysis for both the 2011-TNG and 2021-GTC spectra. See Fig \ref{fig:spec_GTC_and_TNG} left and right panels (also Appendix A). We find that the 2011 TNG pre-CL optical spectrum shows a stronger blue continuum than that of the GTC/OSIRIS spectrum. Nevertheless, the emission line features  mostly look similar in both spectra, except for differences in line strength in most of the emission lines (See Tables \ref{Table:dolores_lines} and \ref{tab:gtc_lines}). We see significant differences in [OIII] doublets. We also detect a HeII emission line in the 2021 GTC spectra which was not present in 2011. Both the pre-CL and post-CL spectra show strong intrinsic host galaxy absorption, which we have carefully and uniformly modeled. This indicates a host stellar population dominated by young stars. We refer the reader to Table \ref{tab:gtc_lines} last column for the changes in the line fluxes relative to the pre-CL state.

 \subsubsection{The broad H$\alpha$ line in 2011 spectrum.}
 
 We detect a weak broad H$\alpha$ emission line in the pre-CL TNG spectrum of 2011, with an FWHM of $2600\pm700\kms$, and line normalization of $(2.10\pm1.25)\times 10^{-15}\funit$ (See Table \ref{Table:dolores_lines} and Figure \ref{fig:spec_LRB_zoom}). We applied a likelihood ratio test and found that this broad emission line component is robustly required by the data. We do not detect any broad emission line in the 2021 GTC spectrum. 
 
 \subsubsection{Diagnostic line ratios}

The line ratios [OIII]/H$\beta$ can be computed only after 
stellar emission subtraction, because we detect H$\beta$ in absorption, which pops up as an emission line after host stellar absorption correction. The line ratios [OIII]/H$\beta$ and [NII]/H$\alpha$ derived after stellar template subtraction are also compatible, within the errors, for the two epochs (see Table \ref{tab:BPTratios}). Note that before stellar template subtraction [NII]/H$\alpha$ is higher in the 2011 spectrum compared to the 2021, which can be explained by the decrease in emission flux due to larger stellar contribution in 2011. The values of the diagnostic line ratios of 1ES 1927+654 lies in the border line between AGN and starburst galaxies \citep{Kewley2006}.  The line ratio H$\alpha$/H$\beta$ is slightly larger than the expected for an AGN (which is $\sim 3.1$), and may indicate some host intrinsic reddening affecting the narrow line region.  Thus the results indicate that 1ES 1927+654 lies in the frontier between Seyfert and Starburst galaxies.


\subsection{Arcsecond and mas-scale Radio Observations}

In the radio regime we were able to cover the pre-flaring state, in October 2013 and March 2014, and the post-flaring activity in late 2018 and 2021. Table \ref{Table:radio} summarizes the low and high-resolution radio observations of 1ES~1927+654 currently available with {\it VLA} and {\it VLBA}. The VLA observations from 1992 and 1998 pre-date the CL event and provide a useful baseline for comparison to the later and higher-resolution VLBI data. The radio spectral index between the C and X band VLA observations is $\alpha_r$=$-$1.1, a relatively steep value but not unusual for radio-quiet AGN \citep[][]{barvainis2005}, especially considering the time baseline and high likelihood of variability.

In all epochs we find an unresolved `point source' component, with a size $<0.5-1\pc$. The three repeated observations at 5 GHz show that the radio emission is variable: it drops four-fold from March 2014 to December 2018, and has increased again in March 2021. This behaviour is roughly concurrent with the X-ray flux (See Figure \ref{fig:xray_uv_alpha_ox} panel 6) and we shall discuss this further in the next section. In both post-flare high spatial resolution radio observations (2018 and 2021) we detect a resolved/extended component which accounts for most of the observed flux. 

The extended emission is not detected in the 1.5 GHz October 2013 EVN observation, likely due to the larger synthesized beam. It is also undetected at the higher-resolution 5 GHz EVN observation in March 2014. We further discuss the interpretation of these results in Sections \ref{subsec:corona} and \ref{subsec:extend-radio}.

\begin{table*}
\centering
  \caption{Semi-contemporaneous X-ray and Radio fluxes (either from 1.5 or 5 GHz VLBI) and the Güdel-Benz Relation.}\label{Table:gb}
  \begin{tabular}{cccccccc} \hline\hline 

X-ray Epoch	& Mean 2-10 keV X-ray flux ($F_X$) & VLBI epoch &  VLBI flux ($F_R$) & $F_R/F_X$ & \\
(MM/YY) & ($\funit{}$) & (MM/YY) & ($\funit{}$) & &  \\
\hline
 $^*$05/11& $3.7\times 10^{-12}$ & 08/13 & $5\times10^{-16}$ & $1.35\times10^{-4}$ \\
 $^*$05/11& $3.7\times 10^{-12}$& 03/14 & $1.8\times10^{-16}$ & $4.8\times10^{-5}$ \\
12/18 & $1.7\times10^{-12}$ & 12/18 & $3.5\times10^{-17}$ & $2.0\times10^{-5}$ \\
03/21 & $4.4\times10^{-12}$ & 03/21 & $6.4\times10^{-17}$ & $1.4\times10^{-5}$ \\
\hline
\end{tabular} 

$^*$ There is no contemporaneous X-ray observations of this source along with VLBI in 2013 and 2014. Hence we used the $2-10\kev$ flux from the 2011 \xmm{} observations by \citet{Gallo2013}.  
\end{table*}



\section{Discussion}\label{sec:discussion}
1ES~1927 has been a unique AGN which is traditionally classified as a true type-II AGN, implying that there has been no detection of broad H$\alpha$ and H$\beta$ emission lines, as well as no line of sight obscuration in the optical, UV or X-rays \citep[][and references therein]{Boller2003,Gallo2013}. The most important point in the unification theory is that all AGN has a BLR, and when we do not detect any BLR it means that our line of sight to the central regions is intercepted by an optically thick torus. This notion has already been challenged by this source by not having any BLR. In an interesting turn of events, this source flared up in UV/optical in Dec 2017 \citep[detected by ATLAS survey][]{trak19} by four magnitudes in just a matter of ~ weeks and a `transient' broad line region (FWHM$\sim 17,000\kms$) appeared only after $\sim 150$ days, which gradually vanished over a period of $\sim$1 year. This proved that the broad line region is present but the central source in its normal state is not luminous enough to light it up. The other interesting highlights from the violent changing look phase are: (1) The UV/optical flare happened in Dec 2017 which gradually decreased. (2) Large Balmer decrement happened in about $\sim 100-200$ days after the flare indicating the presence of dust along the line of sight. (3) The corona completely vanished in Aug 2018 and again came back to normalcy, but did not follow the pattern of the UV/optical flare, implying that the standard disk-corona relations did not hold during the violent event. In this work we investigate the pre- to post-CL state of the central engine using multi-wavelength observations. Below we discuss different science questions in relation to the results found in this work.

\begin{figure}
    \centering
     \includegraphics[width=0.85\linewidth]{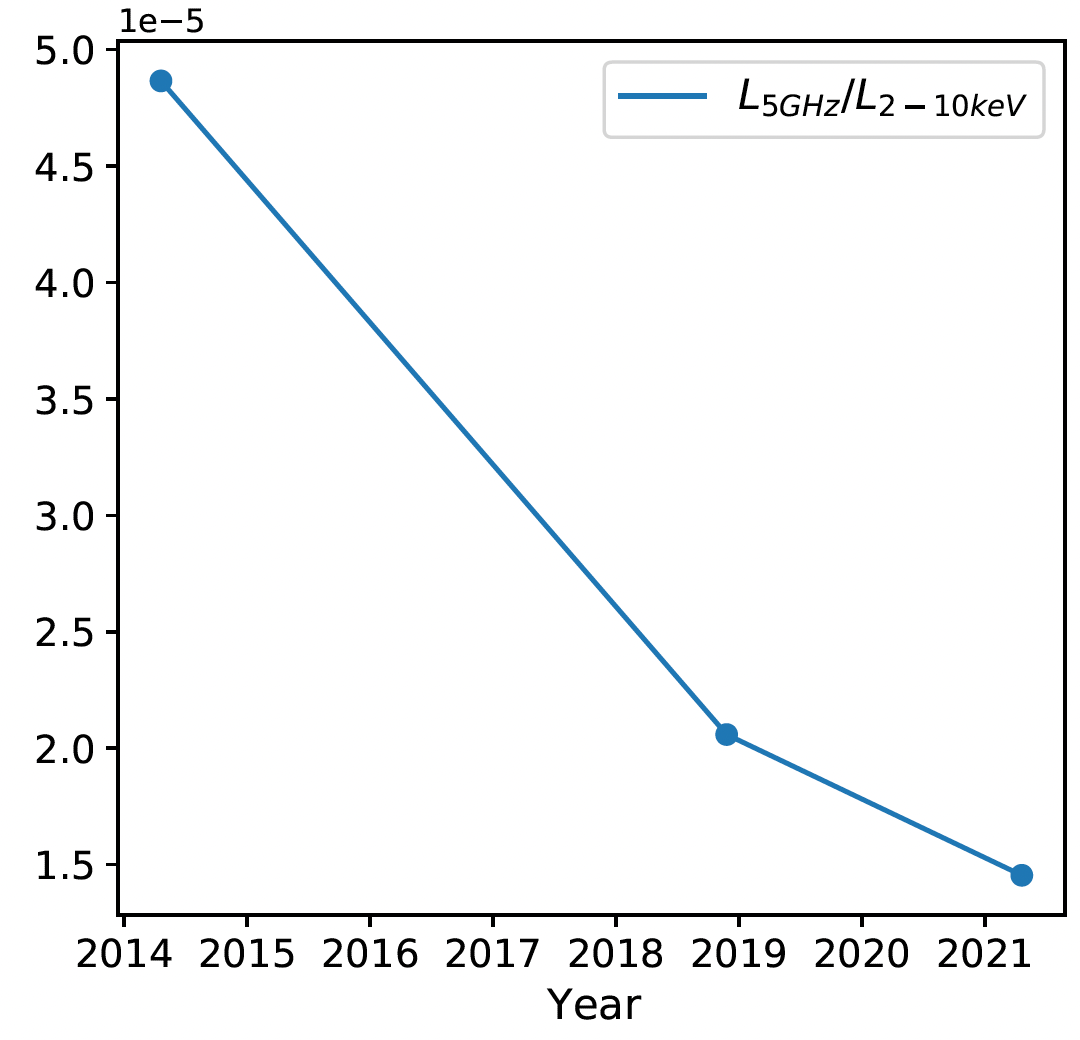}
    \caption{ The light curve of the ratio of the 5GHz mono-chromatic core radio luminosity with the $2-10\kev$ luminosity, popularly known as the Gudel-Benz relation. See Table \ref{Table:gb} for details.}
    \label{fig:Radio-Xray-lightcurve}
\end{figure}


\subsection{What caused this event: Is it a TDE or magnetic flux inversion?}\label{subsec:TDE}

Although the initial papers \citep{trak19,Ricci2020} reporting this unique event suspected a TDE as the source of the CL event, there were already some concerns. For example, \cite{trak19} reported that they did not detect the emission lines in the optical spectra that one would normally detect for a TDE. Similarly \cite{Ricci2020} mentioned that although the UV flux dropped monotonically after the flare, the X-ray flux showed entirely different behavior, which is not what one would expect from a TDE. For the first time in this work we report that the source is back to its normal state and obtain a light curve in UV and X-rays spanning over 3 years. We could fit the UV light curve with an exponential function and found that the slope is much shallower $\propto t^{-0.91\pm0.04}$, pointing to the fact that it could be something other than a TDE which may have triggered the event. We however, cannot rule out the TDE scenario from the shallow slope since it is not clear how a TDE in a pre-existing accretion disk would behave \citep{chan19}. In a sample of 39 TDEs, \cite{van2021} found that the median power-law index is $b=-1.6$, which is consistent with the value $b=-5/3$ which one expects for the full disruption of a star.

In any case, given that (1) the UV slope is $\propto t^{-0.91\pm 0.04}$, (2) the X-ray showed completely uncorrelated behavior in the entire duration of the CL event, (3) the hard X-ray completely vanished after 200 days of the flare, and (4) everything is back to normal in $\sim 1400$ days, we wanted to test an alternative hypothesis, namely the `magnetic flux inversion' scenario presented in \cite{scepi21}, in which the mass accretion rate and the magnetic flux on the black hole are the two independent parameters controlling the evolution of the CL event.

\cite{Ricci2020} suggested that the optical changing look and the X-ray luminosity variations of 1ES~1927+654 can be explained by a TDE which has destroyed the inner accretion disk, and hence the corona. However, we note that the UV luminosity monotonically decreased since the flare, yet the X-ray luminosity was unchanged intially for 100 days, and then it started to fall until it reached a minimum after 200 days of the flare (See Fig \ref{fig:xray_uv_alpha_ox} panels 1 and 2) when the X-ray coronal emission completely vanished for a period of ~2 months (See Table \ref{Table:xray_obs}). The coronal emission revived in October 2018, following which the spectra started to become hard. Thus the optical changing look event and the X-ray luminosity are completely uncorrelated. In most cases of `typical changing look AGN' the X-ray luminosity follows the UV, such as those found in SDSS J015957.64+003310.5 \citep{Lamassa2015}, Mrk~1018 \citep{Husemann2016}, NGC~1566 \citep{parker2019}, and HE~1136-2304 \citep{zetzl2018}. 

\begin{figure*}[h!]
    \centering
    \includegraphics[width=10cm,angle=0]{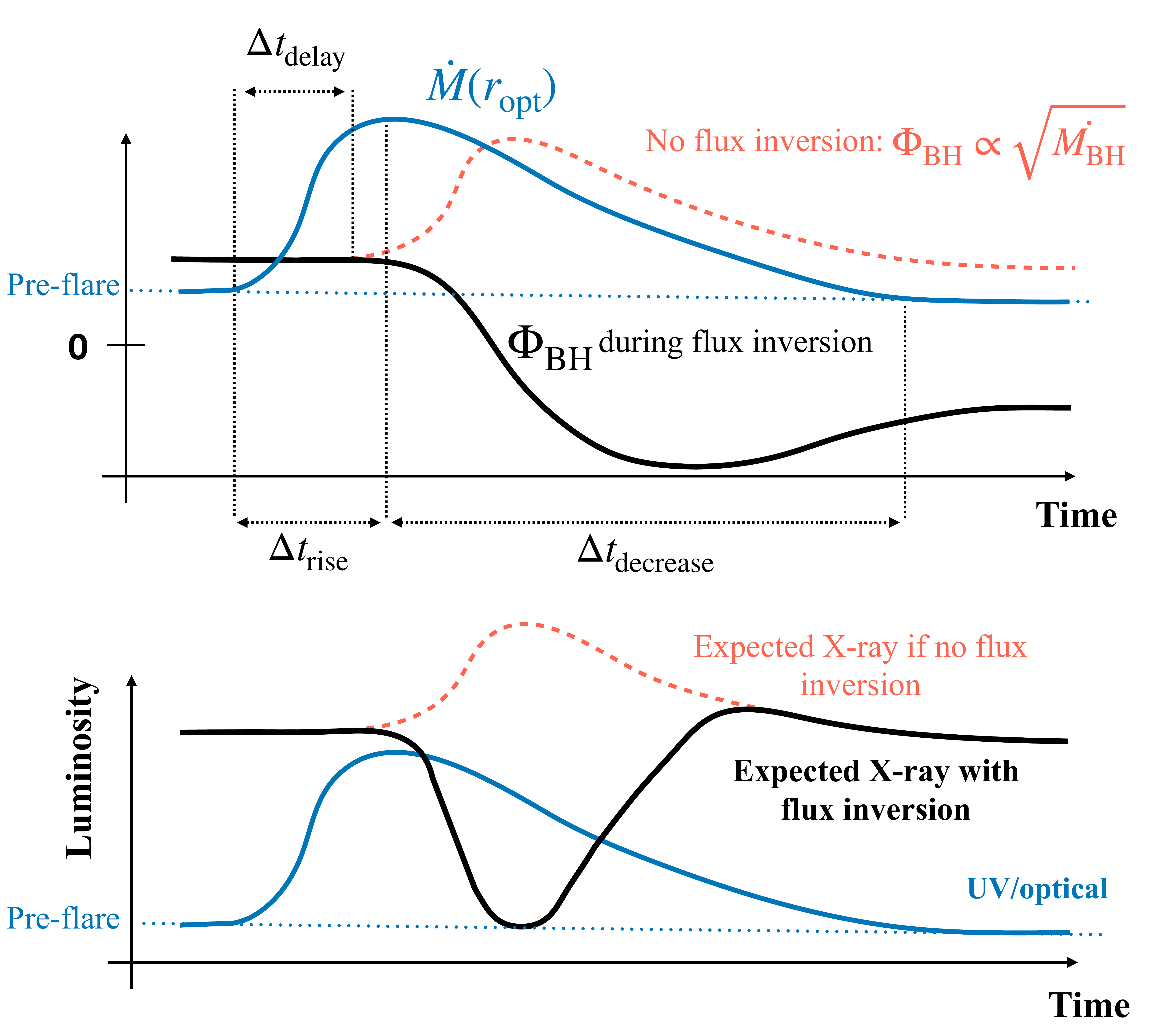}
    \caption{Schematic diagram explaining how a flux inversion event leading to the destruction and regeneration of a magnetically arrested disk (MAD) can explain the separate evolution of the X-rays and the UV of 1ES~1927+654. {\tt Top panel}: Physical parameters controlling the evolution of the CL AGN. The mass accretion rate passing through the disk at the optical-UV emission radius, $\dot{M}(r_{\rm opt})$, is shown as a solid blue line and the magnetic flux on the black hole, $\Phi_{\rm BH}$, in the case of no flux inversion and in the case of a flux inversion are shown respectively as a red dashed line and a solid black line. {\tt Bottom panel:} Dependence of the X-ray and UV luminosities on the physical parameters shown on the top panel. The UV luminosity follows $\dot{M}$ while the X-ray luminosity follows the absolute value of the magnetic flux on the black hole. There is a delay, noted $\Delta t_{\rm delay}$, between the evolution in the UV and the evolution in the X-ray due to the time it takes for a change in the disk to propagate to the black hole. During the flux inversion event the X-ray luminosity goes to zero despite the increase of the accretion rate onto the black hole. When the MAD gets regenerated the X-ray luminosity again follows the trend of the accretion rate onto the black hole, and so of the optical/UV.}
    \label{fig:Mag_flux_inversion}
\end{figure*}

\begin{figure*}[h!]
    \centering
    \includegraphics[width=16cm,height=10cm,angle=0]{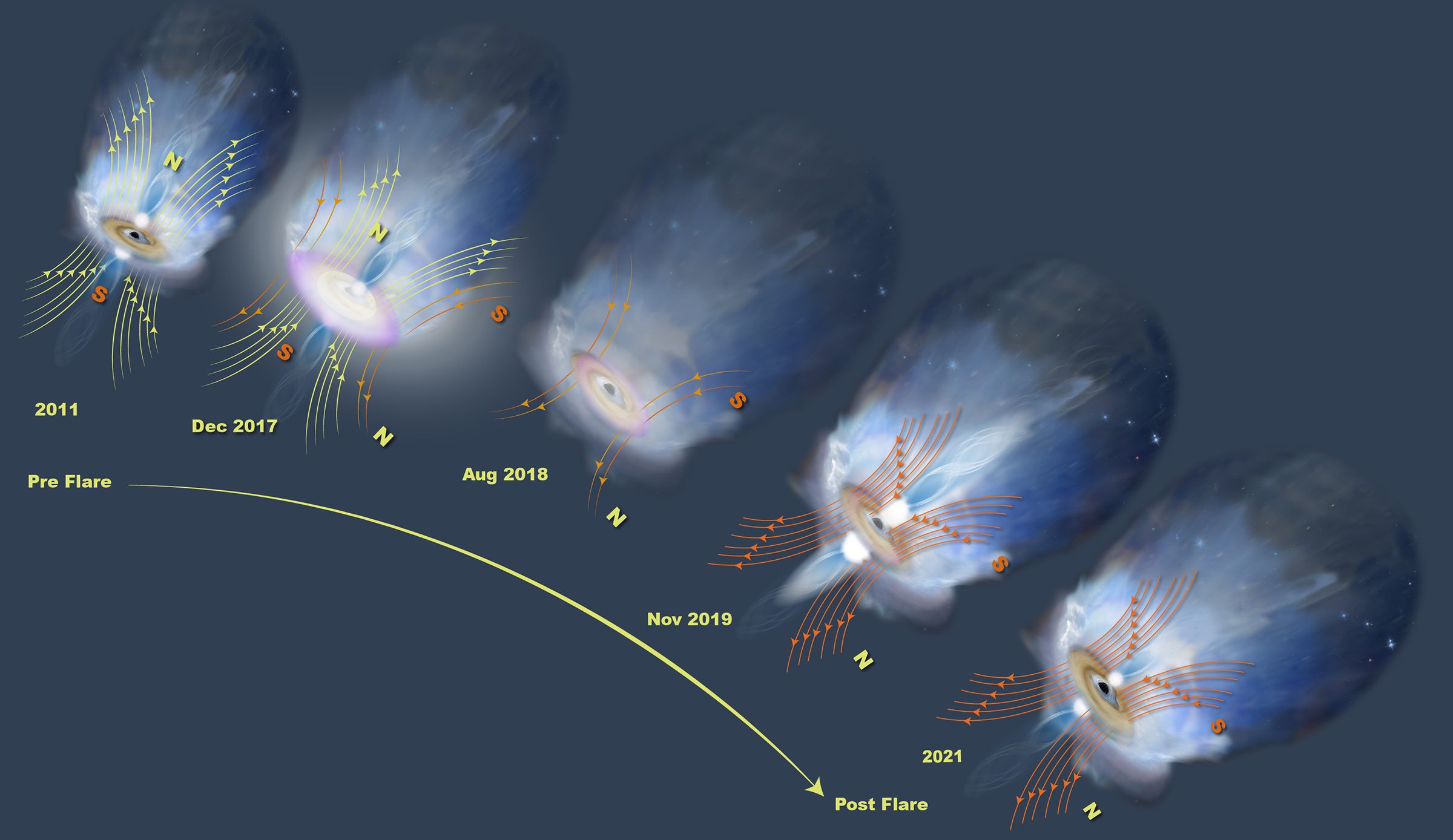}
    \caption{ Cartoon representing the magnetic flux inversion event. From left to right are the preflare to postflare states of the central engine. The initial direction of the magnetic field is depicted by yellow lines, while the reversed polarity is shown by red lines. We discuss the individual panels here, {\it 2011:} The pre-flare state of the AGN where the polarity of the magnetic field threading the accretion disk is in one direction. {\it Dec 2017:} The violent changing look event happens, with the accretion disk brightening up by four magnitudes. This is due to an increase in the accretion rate, possibly related to the change of magnetic polarity in the disk shown in red lines. {\it Aug 2018:} The magnetic flux on the black hole is cancelled by the advection of magnetic flux of opposite polarity and hence the corona vanishes. Note that the accretion disk has dimmed from the 2017 state. {\it Nov 2019:} The magnetic field with reversed polarity now gains strength and hence the corona is revived. Note that the intensity of the corona is several factors larger than in 2011 because the accretion rate is larger than in its pre-flare state value as can be seen by looking at the UV disk emission. The accretion disk still continues to dim. {\it 2021:} The present day scenario, where the magnetic field with reversed polarity has fully formed and the corona and the disk have been restored to their pre-flare 2011 state. Note that the polarity of the magnetic field is opposite that of 2011.
     }\label{fig:Graphics_of_inversion}
\end{figure*}

The uncorrelated evolution of the optical/UV and X-ray suggests that two separate physical parameters are changing during this event. Following \cite{scepi21}, we suggest that the optical/UV are related to a change in the mass accretion rate at some large radii, $\dot{M}(r_\mathrm{opt})$, and that the X-rays come from very close to the black hole and are related to a change in the magnetic flux onto the black hole, $\Phi_\mathrm{BH}$. In a magnetically arrested disk (MAD), $\Phi_\mathrm{BH}$ is proportional to the square root of $\dot{M}_\mathrm{BH}$, the accretion rate on the black hole. Hence, the optical/UV and the X-rays should be correlated but with a delay, $\Delta t_\mathrm{delay}$, corresponding to the time it takes for an increase in the accretion rate in the disk to propagate onto the black hole (see blue solid line and red dashed line on Figure \ref{fig:Mag_flux_inversion}). However, if the magnetic flux brought in by the disk suddenly changed polarity that would destroy this correlation and suddenly shut off the source of X-rays (see black solid line on Figure \ref{fig:Mag_flux_inversion}). Thanks to our new observations we are now able to further constrain this scenario. Figure \ref{fig:Graphics_of_inversion} shows a cartoon depiction of the magnetic flux inversion event. We note below the most important information that can be extracted from these archival and new observations:

\begin{itemize}

\item The fact that the X-ray and UV luminosities go back to their initial values after the whole event suggests that the changing-look event is not triggered by a change in mass supply from the environment. The return to the pre-flare state seems more consistent with an internal mechanism related to the change of polarity. This is reminiscent of \cite{dexter2014}, where the authors observed an increase in $\dot{M}$ during a flux inversion event in their simulation. 

\item In the flux inversion scenario, the X-rays always lag the optical/UV by $\Delta t_\mathrm{delay}$, which corresponds to the time for any change in the disk to reach the black hole (see Figure \ref{fig:Mag_flux_inversion}). We can measure this delay by the time it takes for the X-rays to show a change once the optical/UV changing-look event has started. We estimate that $\Delta t_\mathrm{delay}\approx 150$ days from the data of \cite{trak19,Ricci2020}, and Figure \ref{fig:xray_uv_alpha_ox}.

\item The evolution of the X-rays is complicated because it depends on two parameters, the mass accretion rate $\dot{M}_\mathrm{BH}$ and the magnetic flux $\Phi_\mathrm{BH}$. $\dot{M}_\mathrm{BH}$ should follow the trend of the optical/UV with a delay of $\Delta t_\mathrm{delay}\approx 150$ days as stated above. This means that $\dot{M}_\mathrm{BH}$ and the X-ray luminosity should reach their maximal value around August of 2018. However, this is right during the dip in the X-rays (Fig \ref{fig:xray_uv_alpha_ox}, panel 1). In our scenario, the reason why the X-rays do not reach a maximum in August 2018 is because the magnetic flux on the black hole is cancelled by the advection of magnetic flux of opposite polarity brought along with the increased accretion rate. This destroys the X-ray corona, which is powered by strong magnetic flux near the black hole, and so makes the X-rays drop in spite of $\dot{M}_\mathrm{BH}$ going up. However, after the magnetic flux on the black hole has built up again, recreating the X-ray corona, the X-rays should follow $\dot{M}_\mathrm{BH}$ again. We believe that this happens somewhere around November of 2019 where the X-rays are at their maximum (see Figure \ref{fig:xray_uv_alpha_ox}).

\item After November 2019, the X-rays and the UV are completely correlated again but with a delay of $\approx 150$ days. The fact that the X-rays overshoot their pre-flare value in November 2019 by a factor of $\approx6$ is marginally consistent with the fact that the UV overshoots its pre-flare value by a factor of $\approx 3$ in April 2019 (the closest observation to 150 days before November 2019) as expected in the flux inversion scenario.

\item Again, after November of 2019, the X-rays follow the optical/UV with a lag of $\approx 150$ days. This means that the X-rays should go down to their pre-flare value $\approx 150$ days after the optical/UV. This is consistent with what is observed as the soft X-rays go back to their pre-flare value with a delay of $\approx 200$ days compared to the optical/UV, as can be seen on Figure \ref{fig:xray_uv_alpha_ox}. We note that the duration of the entire event, meaning the time between the departure from pre-flare value and the return to pre-flare value, is the same for the optical/UV and X-rays and is roughly $\approx 1200$ days. This is again consistent with the idea that the changes in the optical/UV and X-ray are due to the same event with, however, the X-rays behaving differently because of their dependence on the magnetic flux on the black hole.

\item The time it takes for the optical/UV to rise is $\Delta t_\mathrm{rise}\approx 200$ days. This time interval corresponds to the time it takes for the inversion event to leave the region of the disk peaking in the optical/UV. It then provides information on the physical size of the inversion event. Moreover, we see that $\Delta t_\mathrm{rise}$ is roughly equal to $\Delta t_\mathrm{delay}$, the time it takes for the inversion event to go from the radius where the disk peaks in the optical/UV to the black hole. This means that the physical size of the inversion event, $\Delta r$, is $\approx r$. In other words, the inversion profile is very gradual. This would explain why such an inversion event could have been preserved in the disk without reconnecting before being able propagate inwards.

\item The radio flux decreases at the same time as the X-ray luminosity drops by 3 orders of magnitude. In \cite{scepi21}, we argued that the X-rays should arise from synchrotron emission of high energy electrons in a corona or failed jet near the black hole. The electrons would also emit synchrotron radiation in radio and so the coincident decrease of the radio and X-rays is consistent with the flux inversion scenario.

 \end{itemize}
 
 


\subsection{The coronal evolution: A radio and X-ray perspective}\label{subsec:corona}



The case of the changing look AGN 1ES~1927+654 serves as a curious test bed to investigate the origin of the coronal emission. This is for the first time in an AGN, the corona completely vanished and then again reappeared in a time scale of $\sim$ 1 year. Similar such phenomenon (contracting corona) has been reported in one of the Galactic black hole binaries \citep{Kara2019}, but never in an AGN. It is evident that the mechanisms that are creating and supplying energy to the X-ray corona must be in a stable equilibrium configuration. That's why it could jump back in a year timescale after being completely destroyed. X-ray and Radio observations provide unique insight into the coronal physics. Variations of a few ten percent of the radio emission are typically observed in AGN \citep{mundell09}. The radio emission in this source is consistent with the one observed in radio-quiet AGN \citep{pan19}. We note the following points in this context:

\begin{itemize}
 \item The central emission radio peak in the AGN 1ES~1927+654 comes from a region $<0.5-1$ pc, while the extended emission covers a region of $4-10\pc$. The observed radio properties are consistent with being produced in the inner region of a low power-jet or a wind or the X-ray emitting corona.

\item A correlation is found between the core radio ($L_R$) and X-ray ($L_X$) luminosities of RQ AGN in the Palomar-Green sample \citep{laor08}, which follow the relation $L_R/L_X\simeq10^{-5}$, remarkably similar to the Güdel-Benz relation for coronally active stars \citep{guedel93}. Although the observed radio cores were on scales of few $100$ parsecs, radio variability constraints on RQ AGN have indicated that most of the emission arises from $\sim$ 10 pc-scale compact regions \citep[e.g.,][]{barv05,mundell09}. This is an indication that the radio emission may originate from the corona. Using the radio and X-ray luminosities of the compact core of 1ES~1927+654, we find that $L_R/L_X\sim10^{-5}$, as illustrated in Table \ref{Table:gb} and Figure \ref{fig:Radio-Xray-lightcurve}. This is an indication that the compact radio emission is related to the region from where the X-ray emission is coming, can possibly be the corona \citep{laor08}.

\item Another interesting aspect is the radio and X-ray light curve. Figure \ref{fig:xray_uv_alpha_ox} panel 6 shows that the core radio flux decreased by a factor of $\sim 4$ in Dec 2018 from its pre-flare state in 2014, just when the X-ray emission was also at its minimum ($\sim $Aug-Oct 2018). The radio flux has since then picked up but not yet reached its pre-flare value. The correlated decrease and then increase of the radio flux with the X-ray flux is likely a sign of the coronal recovery and thereafter a coronal (Synchrotron) origin of the radio emission. 

\item   The radio brightness temperatures of the AGN 1ES~1927+654 $T_B\gtrsim10^6$ K in the pre-flare and the post-flare states (See Table \ref{Table:radio} last column) points to a non-thermal origin of the core radio emission. This could either arise from the X-ray corona or a jet knot \citep{panessa22}.

\item The steep radio spectrum ($\alpha_r=-1.1$) is consistent with an optically thick regime. In the case of an optically thick Synchrotron source, as per the equation 19 in \cite{laor08}, the estimated emission region would be $\sim 0.001\pc$, which is again consistent with the X-ray corona, or a very nascent jet.

\end{itemize}

Given all the evidences we can say that the core radio emission arise from the inner most regions of the AGN and can possibly be related to the X-ray corona or a nascent unresolved jet.


 
 \subsection{The origin of the Soft X-ray emission}
 
 The origin of the soft X-ray emission in AGN is still debated. Popular theories suggest that it could arise out of thermal Comptonization of the UV seed photons by a warm corona \citep{Done2012} or it can also arise out of the reflection of the hard X-ray photons from the corona off the accretion disk \citep{Garcia2014}. Even after considerable studies on the subject \citep{laha2011Mrk704,laha2013IRAS,laha2014ESO198,ehler2018,laha2018ucsd,garcia2019,waddell2019,laha2019,tripathi2019,Ghosh2020,ursini2020,Laha2021} there has not been a consensus on the physics of the origin of the soft X-ray excess. X-ray reverberation lead/lag studies too have yielded conflicting results for different sources \citep{Kara2016}. Hence, the special cases of changing look AGNs give us a unique platform to investigate the origin of this feature, which is ubiquitously present in bright AGN. For example in the CL-AGN Mrk~590, it has been seen that the soft X-ray flux remains bright even when the power law flux goes through a dim state \citep{Mathur2018}, indicating that possibly the soft excess in this source may not arise out of disk reflection of the hard X-ray photons.
 
 The source 1ES~1927+654 has exhibited a strong soft-excess at energies below $\sim 2\kev$ in its pre-CL state in 2011 which could be well modeled by a black body of $\rm kT=0.20\pm 0.01\kev$ (See Table \ref{Table:xray_obs}). \cite{Ricci2021} modeled the \xmm{} X-ray spectra with complex disk reflection models, but could not obtain a reasonable fit. From Figure \ref{fig:soft_hard_correlation} we note that the soft and the hard X-ray show no correlated behavior, whether during the X-ray flare (red points) or in the post-flare states (blue points). \cite{Ricci2021} using high cadence \nicer{} observations reports that at around $\sim 200$ days after the UV flare, the hard X-ray completely vanished (Aug 2018) for around $\sim 3$ months until it re-emerged in Oct 2018. During this time, the soft X-ray was still present, but in a low state. From these evidences we can rule out disk-reflection origin of the soft excess for this source, because in the reflection scenario, the hard X-ray emitting power law is the primary emitter.  
 
 Also, there is no correlation between the UV flux and the soft X-ray flux during and after the changing look event, implying a possibility that the origin of the soft excess may not be from the warm Comptonization of the disk UV photons. However, as a caveat we must note that during the flaring event, the disk was possibly destroyed and hence we do not expect standard models to hold. 

 The other interesting observational points to note in this context for this source are:
 \begin{itemize}
 \item There is no detection of narrow or broad FeK$\alpha$ emission line at $6.4\kev$ \citep{Gallo2013}, which is otherwise commonly detected in most AGN. This indicates a complex reflection geometry of the central engine.

 \item  There is a detection of a narrow emission line at $1\kev$ in the \xmm{} spectra in the post-flare scenario \citep{Ricci2021}. This emission line was not present in 2011 \xmm{} spectrum.

 \item The hardness ratio (HR) plot shows weird variability pattern. See Fig. \ref{fig:xray_uv_alpha_ox} panel 3. During the hard X-ray dip ($\sim 200$ days after the onset of the flare), the HR is very low (coincident with the vanishing of the corona) and it jumps back in Oct 2018, goes down again, and then gradually reaches the pre-CL state over a period of $\sim 300$ days. We note that even though the HR has reached its pre-CL value, both the soft and the hard X-ray fluxes were almost $\sim 8$ times higher than their original pre-CL values (See Fig. \ref{fig:xray_uv_alpha_ox}). More intriguingly, in Feb 2021 when we started monitoring the source with \swift{}, the HR was again as low as the lowest HR state, and then again climbed up to its pre-CL value over a period of $\sim 200$ days. These point towards a complex relation between the emitters of soft and hard X-rays. A time dependent physical modeling can reveal the nature of the soft X-ray emitter.
 
 \item The hard X-ray flux jumps back in Oct 2018 after staying undetected for $\sim 3$ months (from July-Oct 2018), and at the same time, there is a factor of 2 decrease in the UV luminosity, quite distinct deviation from its normal $t^{-0.91}$ exponential drop. See Fig \ref{fig:xray_uv_alpha_ox} panels 2, 3 and 4. It could be that the inner accretion disk empties itself to pump matter for the formation of X-ray corona.

 \end{itemize}


\subsection{Is the disk-corona relation restored?}

Although there is no consensus about the exact origin, geometry and location of AGN corona, there is substantial evidence that the accretion disk and the corona of AGN are coupled to each other over a wide range of mass, luminosity and accretion states, and this coupling is universal across all redshift \citep{lusso2010,lusso2016}. The universal coupling of the UV emitting disk and the corona suggests that the disk supplies the seed photons which the corona upscatters (Inverse Comptonize) and we observe a power law at energies $>2\kev$. The relation between the disk UV photons and the X-ray emission is measured by the correlation between $\alpha_{\rm OX}$ and $L_{\rm 2500 \angs}$ which is very tight even when considering local Seyferts and high redshift luminous quasars \citep{laha2014WAX,Laha2018,Martocchia2017,laha2021natAs}. However, in the case of 1ES~1927 we do not detect any correlation between the UV and the X-ray photons during the violent event. The corona was disrupted much later ($\sim 200$ days) than the first UV flare happened. The $\alpha_{\rm OX}$ values during the flare are reported in Table \ref{Table:xray_obs}. Figure \ref{fig:alphaOX} shows that in the pre-CL state (denoted by red star), the AGN was slightly below the disk-corona relation, but the relation was completely disrupted during the flare, with the parameter values spanning large regions in the phase space. However, as the source returned to the quiescent state (the blue triangles in the lower left corner of Figure \ref{fig:alphaOX}), the $\alpha_{\rm OX}$ values ($\sim 1.02$) returned back to their pre-flare values, indicating that the disk-corona link has again been established. The post-flare values of the $\alpha_{\rm OX}$ and $L_{\rm 2500 \angs}$ lie on the correlation detected in AGN samples \citep{lusso2010,lusso2016}.\\

\subsection{Variability in the narrow line region}

In both the pre-CL and post-CL spectra we detect mostly narrow emission lines (See Tables \ref{Table:dolores_lines} and \ref{tab:gtc_lines}). Except for the H$\beta$ emission line, we find that all the lines have evolved between 2011 and 2021. For example, new emission lines of [OII]3727$\angs$,  HeII4686$\angs$, and OI6300$\angs$ emerged in the post-CL 2021 GTC spectra (See Table \ref{tab:gtc_lines} last column). On the other hand, the [OIII]4959$\angs$ and [OIII]5007$\angs$ doublet became weaker in 2021. The [NII]6548$\angs$ and [NII]6584$\angs$ doublets became stronger. In addition, we also find a new emerging narrow HeII emission line in 2021. If the narrow line region is $\sim 1-10 \pc$ away from the central engine, any flux variations from the AGN would require $\sim 3-30$ years to reach the NLR. May be we have just started to see the changes, given the fact that the 2021 observation was made $\sim 3$ years after the start of the flare. Future monitoring are crucial to detect any further changes, and thereby determine the distance and extent of the NLR.



\subsection{The nature of the broad line region: Does true type-II exist?}

The missing BLR in the AGN 1ES~1927+654 throws a real challenge to unification theory. This source has been classified as true type-II \citep{Gallo2013}, which we know now, may not be true, given the fact that we indeed detect a BLR during the high flux state. The distance of the BLR ($\sim 100-150$ light-days) are also of similar order as that predicted by reverberation mapping studies \cite{Peterson1993}. In our work, we detected a weak broad H$\alpha$ emission line with FWHM$\sim 2644\pm700\kms$ in the 2011 TNG optical spectrum, indicating the presence of a BLR which is very weak. We note that this emission line may not arise from the same BLR region as detected by \cite{trak19}, with a FWHM$\sim17,000\kms$. Interestingly, we do not detect any such broad line in the post-CL 2021 GTC optical spectra. The detection of the weak broad H$\alpha$ emission line poses interesting challenges to our knowledge of true type-II sources.
Perhaps the BLR always existed, it is just that its not bright enough to be detected, given that the source is accreting at a very low rate.

For example, the nearby AGN NGC~3147 has been referred as one of the best cases of a true type-II AGN, implying no absorption along the line of sight, yet no broad line present. However, for a low luminosity AGN like NGC~3147 one would expect to have a broad line which is highly compact and hence hard to detect against the background host galaxy. \cite{bianchi2019} carried out a  Narrow (0.1 arcsec $\times$ 0.1 arcsec) slit Hubble Space Telescope (HST) spectroscopy for this source, which allowed them to exclude most of the host galaxy light. Very interestingly they detected a broad H$\alpha$ emission line with a full width at zero intensity (FWZI) of $\sim 27,000\kms$. This result challenges the very notion of `` true type-II classification". Coupling these with our results, we propose that there may not be any source which can be called a true type-II. Future narrow slit spectroscopy of low accretion sources can reveal further information.


\subsection{What is the extended radio emission? Is there any evidence of jet formation?}\label{subsec:extend-radio}

Archival, or pre-flare, VLA observations of the source at 1.4 GHz (1995 NVSS, \citealt{nvss}), 5 GHz (1992) and 8 GHz (1998) show unresolved cores ($\sim$ tens of pc to kpc) with peak flux of 40, 16, and 9 mJy respectively. This can be compared with VLBI pre-flare observations, where we see that the total flux is only one quarter to one half of these values, similar to the reductions in observed flux on arcsecond to mas scales for radio-loud AGN (e.g., \citealt{giov05}). The extended emission detected in the VLBI imaging has very similar flux and shape through December 2018 and March 2021. It is unclear if the absence of extended emission in the matching-resolution March 2014 epoch is intrinsic or mainly undetectable due to non-optimal $u-v$ coverage and depth of the observation. The comparison of the NVSS and EVN fluxes at 1.5 GHz do suggest extended emission is present, but the resolution of NVSS is very \textcolor{blue} {poor} (45$''$), and we also see discrepancies between the total VLBI-scale flux in 2018/2021 and the corresponding VLA observations. Whether this is due to variability or additional extended emission beyond the VLBI scale, we cannot say. 

Extended radio emission from RQ AGN may arise in different ways \citep[see e.g.][for a recent review]{pan19}, with the spatial extent being an important clue. If on the scale of the galaxy it may be attributed to star formation, which is verifiable with far-IR observations.  If the structure is small enough and not resolved with the VLA, but resolved on the mas scale it could be a nascent jet, as was recently discovered in the CL AGN Mkn 590 \cite{Mrk590jet}. In the case of 1ES~1927, the best fit model is an extended uniformly illuminated disk and the emission is seemingly isotropic. This is contrary to the general picture of jets from black holes, which are usually highly collimated. An extended disk wind bright due to free-free emission may also be a possibility \citep{pan19}, but we need higher frequency observations and better constraints on spectral slope to verify this.
Further, naively using the approximate disk size of 5 light years and assuming it grew to this size in $\sim$ 1 year from the onset of optical-UV flaring due to the TDE, the apparent speed $\gtrsim2c$.  This is obviously problematic for either a collimated outflow (which our observations due not resemble) or a sub-relativistic disk wind, or essentially any origin which is novel since the CL event. The extended emission may also possibly represent the outskirts of a maser disk, as found in NGC 1068 or NGC 4258 (e.g., \citealt{greenhill95,gallimore04}), which can only be verified through deep spectroscopic studies.


\section{Conclusions}\label{sec:conclusions}

In this work we report the evolution of the radio, optical, UV and X-rays from the pre-flare state through mid-2021 with new and archival data from the {\it Very Long Baseline Array}, the {\it Very Large Array}, the {\it Telescopio Nazionale Galileo}, {\it Gran Telescopio Canarias}, {\it The Neil Gehrels Swift observatory} and \xmm{}. The main results from our work are:

\begin{itemize}

\item The source has returned to its pre-CL state in optical, UV, and X-ray; the disk--corona relation has been re-established as has been in the pre-CL state, with an $\alpha_{\rm OX}\sim 1.02$. 

\item Since the central engine returned back to its pre-CL state in a matter of a few years, we conjecture that something internal to the accretion disk triggered this event, and not something external (such as  a change in the external matter supply). We conjecture that a magnetic flux inversion event is the possible cause for this enigmatic event. 

\item  The evolution of the optical/UV emission and the X-ray emission can be well explained in a scenario where both emissions depend on the underlying accretion rate. The apparent non-correlation of the two bands stems from:  1) that the X-rays depend on the accretion rate near the black hole and the optical/UV depend on the accretion rate further away in the disk and 2) that the X-rays also depend on the magnetic flux that can be inverted during the event. 

\item  The UV light curve follows a shallower slope of $\propto t^{-0.91\pm0.04}$ compared to that predicted by a tidal disruption event, indicating a possibility of a non-TDE event. 

\item  In the optical spectra, we mostly detect narrow emission lines both in the pre-CL (2011) and post-CL (2021) spectra. Perhaps the BLR in this source always existed, it is just that it's not bright enough to be detected, given that the source is accreting at a very low rate.


\item  We also detect substantial line-of-sight host-galaxy stellar absorption at both epochs, and variability in the narrow emission line fluxes at the two epochs. 

\item  The compact radio emission which we tracked in the pre-CL (2014), during CL (2018) and post-CL(2021) at spatial scales $<5\pc$ was at its lowest level during the changing look event in 2018, nearly contemporaneous with a low $2-10\kev$ emission. This points to a Synchrotron origin of the radio emission from the X-ray corona. 

\item The radio to X-ray ratio of the compact source $L_{\rm Radio}/L_{\rm X-ray}\sim 10^{-5.5}$, follows the Gudel-Benz relation, typically found in coronally active stars, and several AGN. 

\item We also detected extended radio emission around the compact core at spatial scales of $\sim 4-10 \pc$, both in the pre-CL and post-CL spectra. We however, do not detect any presence of nascent jets.

\end{itemize}

\clearpage{}
\section{Acknowledgements}

The authors acknowledge useful discussions with A. Vazdekis.
The material is based upon work supported by NASA under award number 80GSFC21M0002.
JBG and JAP acknowledges financial support from the Spanish Ministry of Science and Innovation (MICINN) through the Spanish State Research Agency, under Severo Ochoa Program 2020-2023 (CEX2019-000920-S). SB acknowledges financial support from the Italian Space Agency (grant 2017-12-H.0). SL is thankful to the \swift{} team for granting the director's discretionary time to observe the source at a regular cadence. SL is thankful to Jay Friedlander who has created the graphics of the magnetic flux inversion event (Fig \ref{fig:Graphics_of_inversion}). MN is supported by the European Research Council (ERC) under the European Union’s Horizon 2020 research and innovation programme (grant agreement No. 948381) and by a Fellowship from the Alan Turing Institute. E.B. acknowledges support by a Center of Excellence of THE ISRAEL SCIENCE FOUNDATION (grant No. 2752/19).

\section{Data availability:}

For this work we use observations performed with the GTC telescope, in the Spanish Observatorio del Roque de los Muchachos of the Instituto de Astrof\'isica de Canarias, under Director\textquotesingle s Discretionary Time (proposal code GTC2021-176, PI: J. Becerra).

This research has made use of new and archival data of \swift{} observatory through the High Energy Astrophysics Science Archive Research Center Online Service, provided by the NASA Goddard Space Flight Center.


\bibliographystyle{aasjournal}
\bibliography{mybib}

\clearpage

\appendix{}

\counterwithin{figure}{section}
\counterwithin{table}{section}

\section{Optical Spectra}

\begin{figure*}
    \centering
    \includegraphics[width=\linewidth]{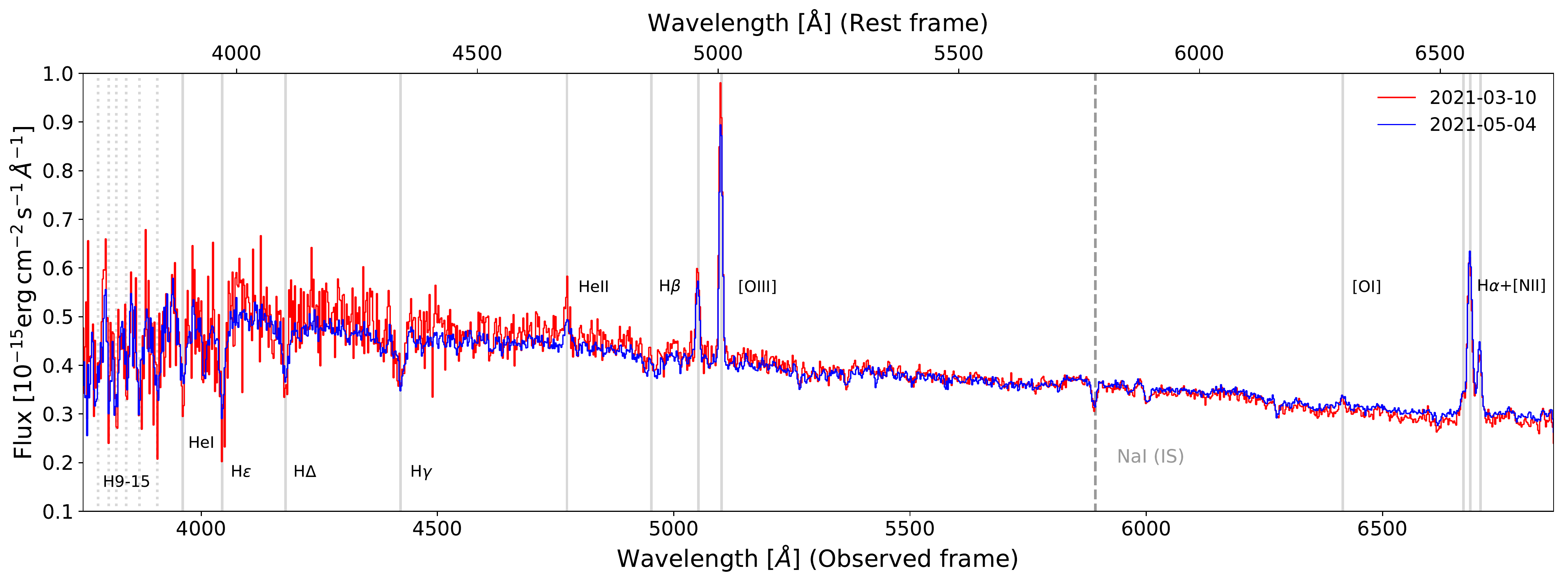}
    \caption{The post-changing-look optical spectra of 1ES~1927+654 taken with GTC in March and May 2021 (red and blue respectively). The grey bands denote the different emission and absorption features. The spectra has not been corrected for host galaxy absorption}
    \label{fig:spec_GTC_nostellar}
\end{figure*}

\begin{figure*}
    \centering
      \includegraphics[width=12cm]{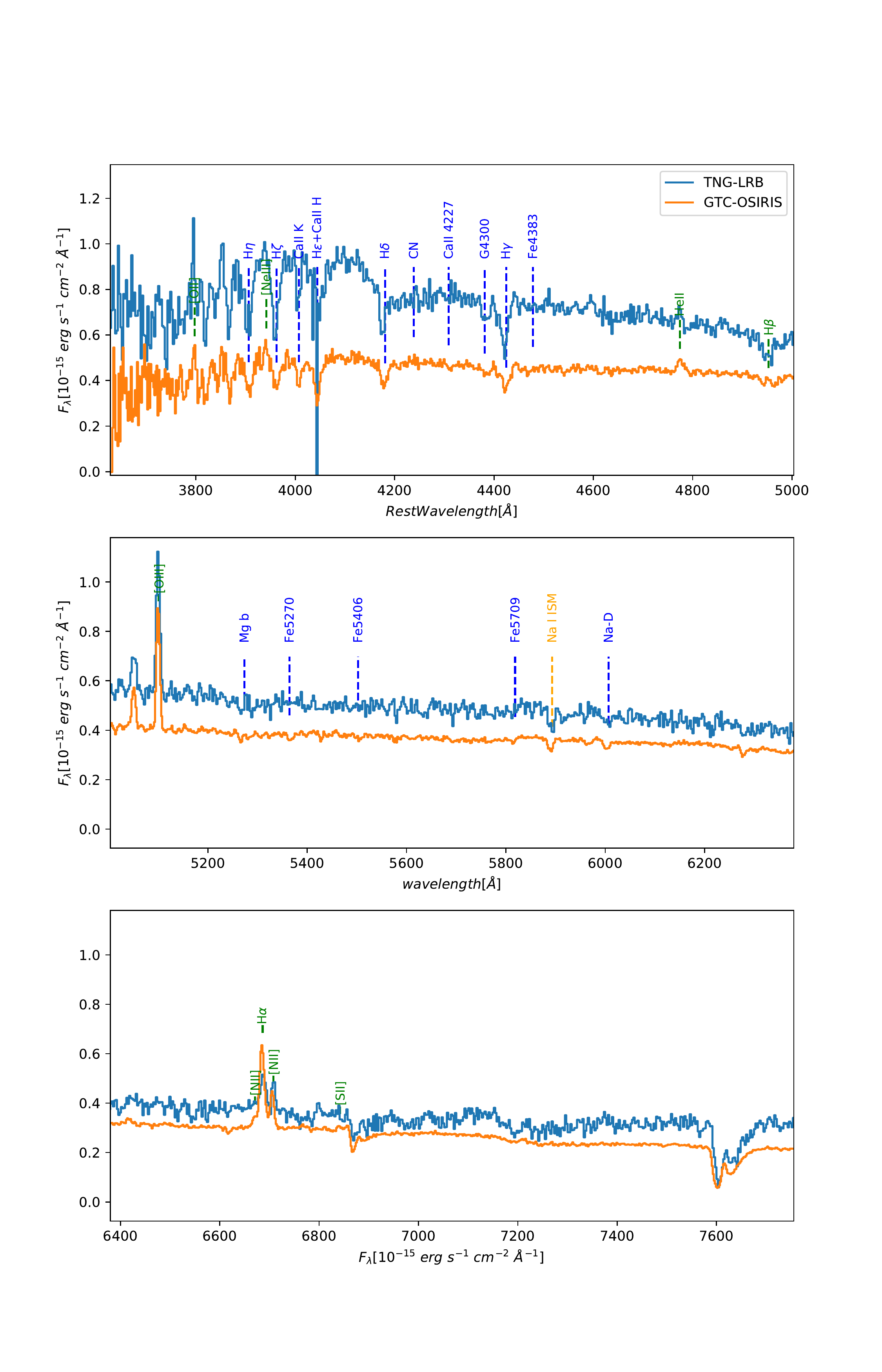}
    \caption{The comparison between the pre-CL and post-CL optical spectra. The blue spectra is the 2011 TNG observation, while the orange one is from 2021 GTC observations. The 2011 spcetrum is much bluer than the 2021.}
    \label{fig:optical-comparison}
\end{figure*}

\begin{figure*}
    \centering
    \includegraphics[scale=0.45]{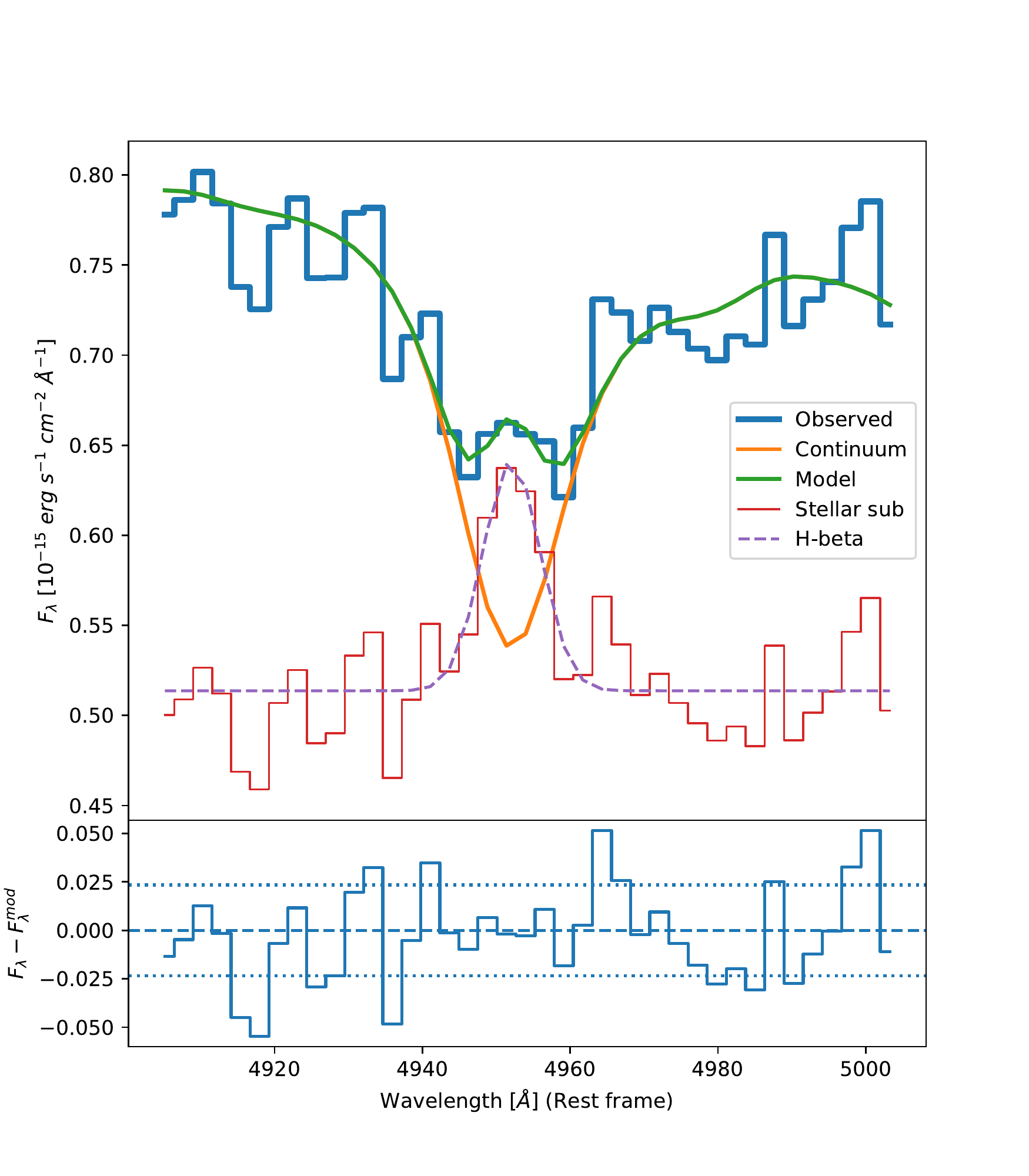}
    \includegraphics[scale=0.45]{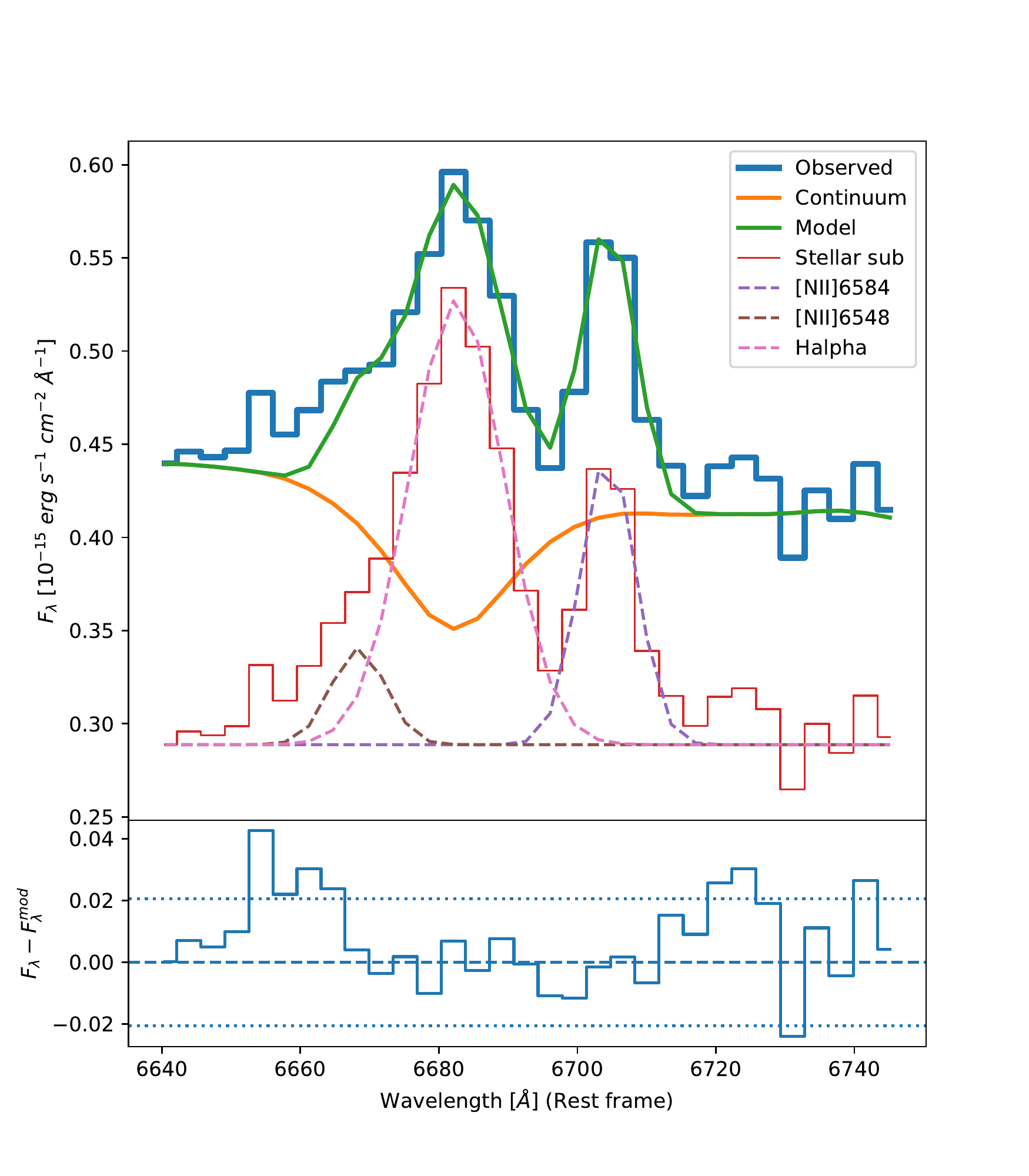}
    \includegraphics[scale=0.45]{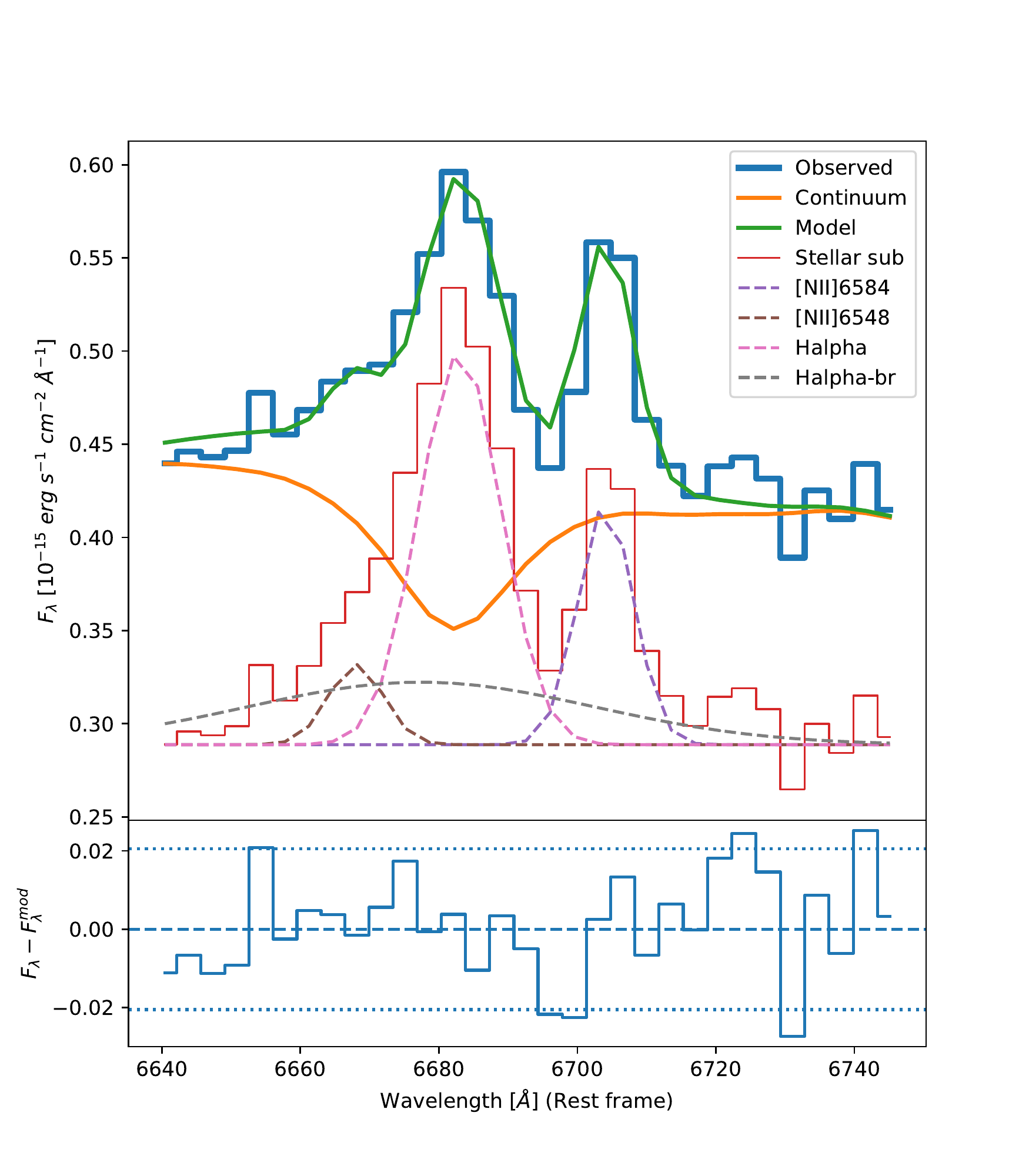}
    \caption{The 2011 TNG Optical spectrum of 1ES~1927+654 zooming in the crucial H$\beta$ and H$\alpha$ regions. For all the panels, the stellar model and the emission line fits are shown. The residuals compared to the full model, including the stellar model plus the emission line fits, are plotted in the bottom panel of each figure. {\it Top Left:} The H$\beta$ line detected in absorption in the observed spectrum, and in emission after correcting for the host stellar absorption. {\it Top Right:} Same as left figure, but for the H$\alpha$ emission line complex. {\it Bottom:} Same as top right figure, but after fitting with a broad Gaussian to model the broad H$\alpha$ emission line.}
    \label{fig:spec_LRB_zoom}
\end{figure*}


\begin{figure*}
    \centering
    \includegraphics[scale=0.3]{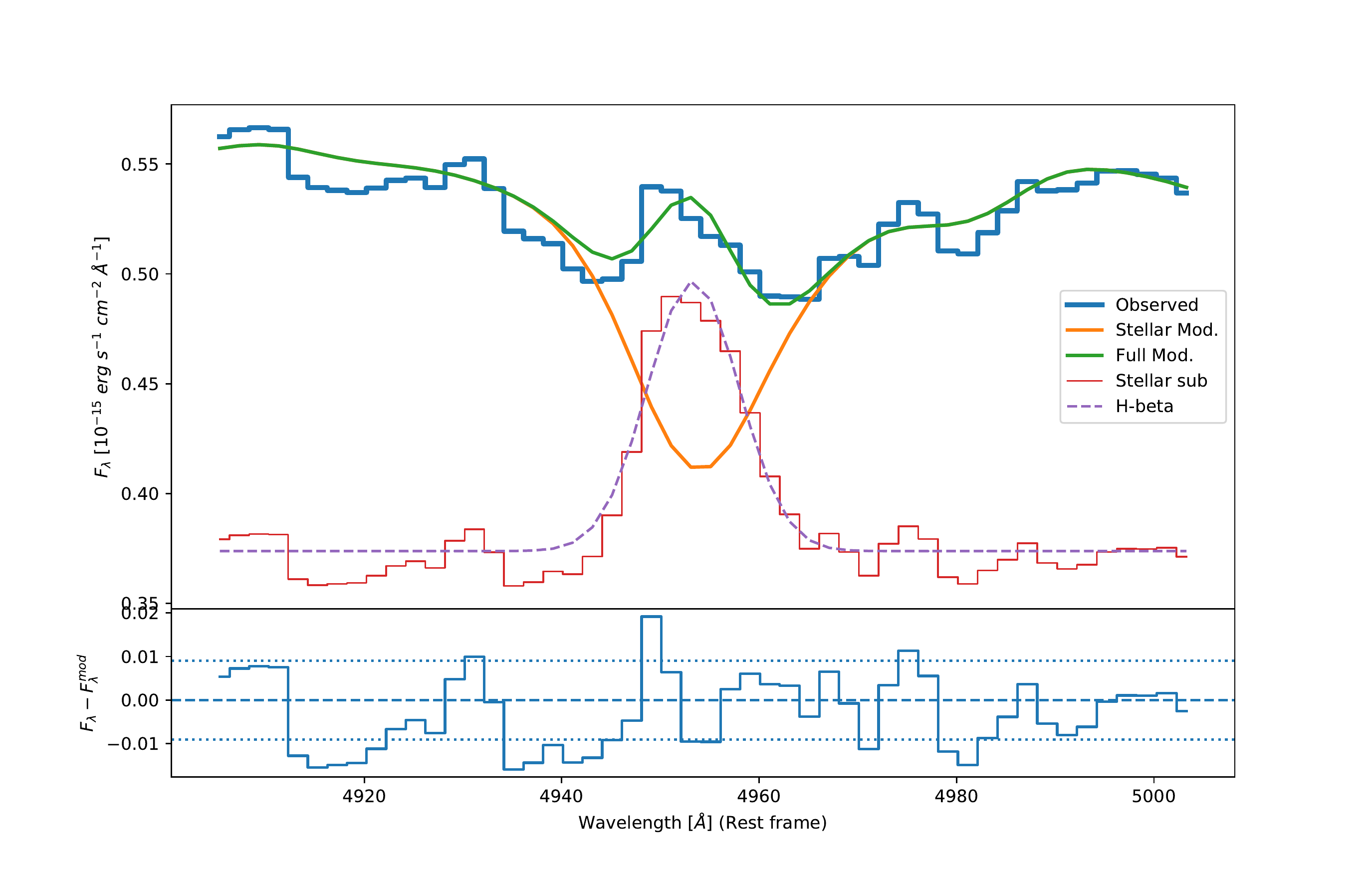}
    \includegraphics[scale=0.3]{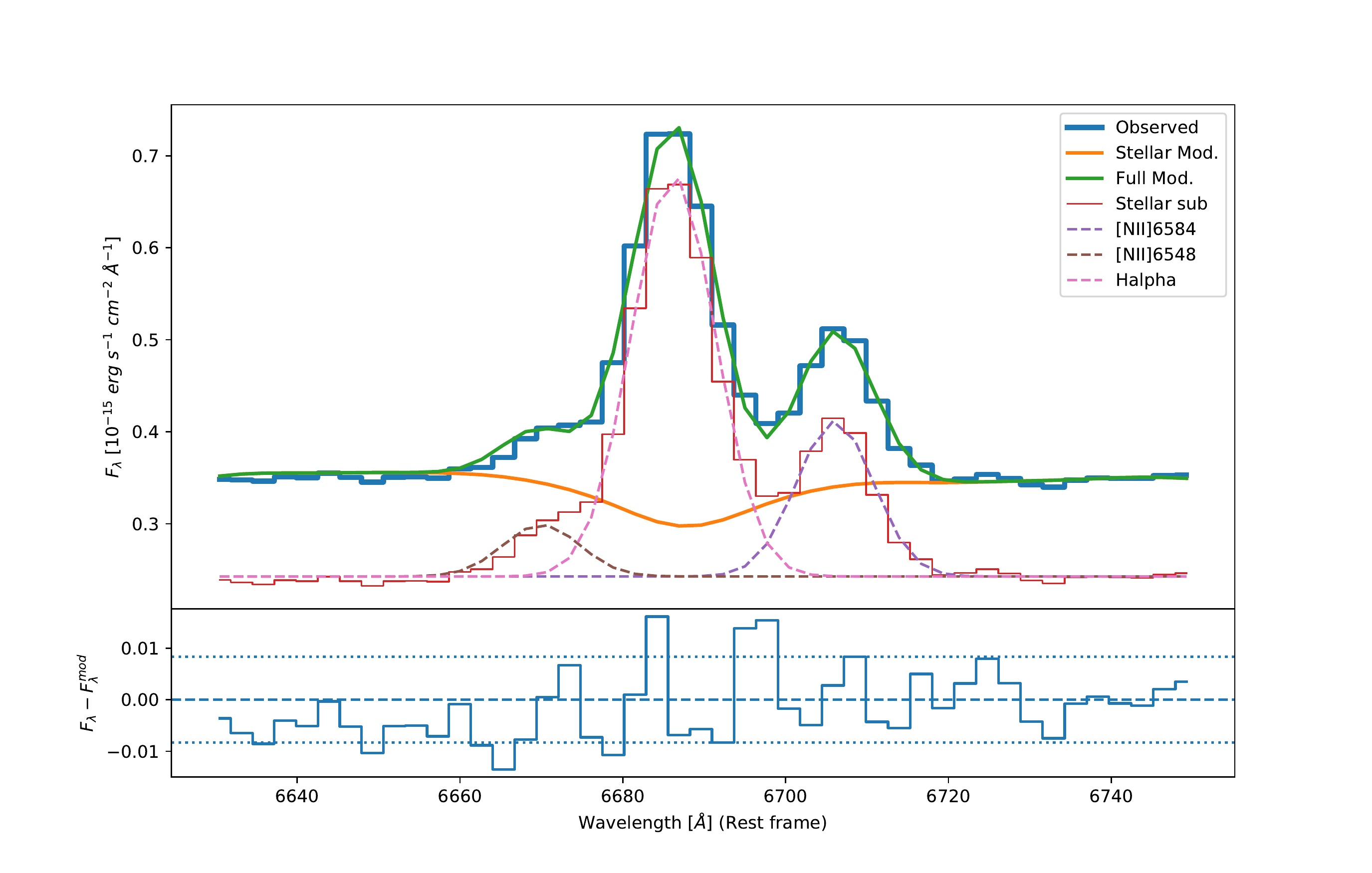}
    \caption{The 2021 GTC Optical spectrum of 1ES~1927+654 zooming in the crucial H$\beta$ and H$\alpha$ regions. The stellar model and the emission line fits are shown in the top panel of each of the two figures. The residuals compared to the full model, including the stellar model plus the emission line fits, are plotted in the bottom panel. We do not detect any broad emission line in the 2021 spectra.}
    \label{fig:spec_GTC_zoom}
\end{figure*}

\subsection{The Non--parametric approach to fitting the optical emission lines}

First, we have measured the spectral features using a non--parametric mode, after continuum subtraction. The continuum is subtracted by fitting a straight line using regions adjacent to the spectral feature to be measured. The line center is estimated as the mass-center, the flux by adding the emission within the line interval and the EW as the flux divided by the continuum at the line center.  
Most of the emission lines seems to have compatible values for the flux and EWs within $2\sigma$. The only exceptions are the complex $H\alpha+[NII]$ which shows larger emission flux in the 2021 spectrum. 
Despite the difficulties to measure the absorption features in the TNG/LRB spectrum due to the low S/N ratio in the blue part, the EWs of absorption features are compatible between the 2 epochs, specially the Balmer lines $H\delta$ and $H\gamma$. The exception is the line $H\beta$, which shows a smaller EW in the 2021 spectrum. This is compatible with a deeper absorption contribution (in absorbed flux but not in EW) or alternatively an increase in the emission contribution (the H$\beta$ profile is formed by the contributions of the stellar absorption and the emission one). Later, after modelling stellar population emission, one could discern which scenario is favoured. High order Balmer lines are more dominated by the stellar emission and low order H$\alpha$ and H$\beta$ are more contaminated by the emission from the ionized gas of the NLR.   
From this comparison, it can be said that the stellar emission seems to be compatible in the spectra taken in the two epochs, 2011 and 2021, despite the difference in flux which can be due to the different aperture used.   
In Figure \ref{fig:spec_GTC_nostellar} we show the two GTC spectra obtained in March and May 2021. In Figure \ref{fig:optical-comparison} we show the comparison between the two spectra at the two epochs: 2011 TNG and 2021 GTC. 

\subsection{Zooming into the Balmer emission line profiles}

In figures \ref{fig:spec_LRB_zoom} and \ref{fig:spec_GTC_zoom} we show the zoomed in data+model+residuals of the H$\alpha$ and H$\beta$ complex. We did not detect any broad emission line consistent with that observed by \citet{trak19}. See Section \ref{subsubsec:broad_lines} for details.


\end{document}